\documentclass[11pt]{article}
\usepackage[utf8]{inputenc}
\usepackage{amsmath}
\usepackage{amsfonts}
\usepackage{graphicx,color}
\usepackage{hyperref}
\usepackage{multirow}
\usepackage[dvipsnames]{xcolor}

\usepackage{listings}
\usepackage{color}

\usepackage{algorithm}
\usepackage{algorithmic}

\newcommand{\code}[1]{{\normalfont\ttfamily\hyphenchar\font=-1 #1}}
\let\proglang=\textsf
\newcommand{\pkg}[1]{{\fontseries{m}\fontseries{b}\selectfont #1}}

\hypersetup{
  colorlinks=TRUE,
  linkcolor=blue,
  urlcolor=blue,
  citecolor=blue
}
\usepackage{authblk}
\usepackage{epsfig}
\usepackage{amssymb}
\usepackage{fancyhdr}
\usepackage{amsmath}
\usepackage[round,authoryear,sort&compress]{natbib}
\usepackage{caption}
\usepackage{geometry}
\title{\pkg{hhsmm}: An \proglang{R} package for hidden hybrid Markov/semi-Markov models}
\author{Morteza Amini\thanks{morteza.amini@ut.ac.ir}, ~
Afarin Bayat\thanks{aftbayat@gmail.com} ~  and~
Reza Salehian \thanks{reza.salehian@ut.ac.ir} \\
{\small Department of Statistics, School of Mathematics, Statistics and Computer Science, University of Tehran, Tehran, Iran}}

\begin{document}
\maketitle

\begin{abstract}
This paper introduces the \pkg{hhsmm} \proglang{R} package, which involves functions for initializing, fitting, and predication of  hidden hybrid Markov/semi-Markov models. These models are flexible models with both Markovian and semi-Markovian states, which are applied to situations where the model involves absorbing or macro-states. The left-to-right models and the models with series/parallel networks of states  are two models with Markovian and semi-Markovian states.  The \pkg{hhsmm} also includes { Markov/semi-Markov switching regression model as well as the auto-regressive HHSMM, the nonparametric estimation of the emission distribution using penalized B-splines, prediction of future states and } the residual useful lifetime estimation in the \code{predict} function. The commercial modular aero-propulsion system simulation (C-MAPSS) data-set is also included in the package, which is used for illustration of the application of the package features. The application of the \pkg{hhsmm} package to the analysis and prediction of the Spain's energy demand is also presented.  
\end{abstract}

\noindent{\bf Keywords:} {Continuous time sojourn, EM algorithm, Forward-backward, Mixture of multivariate normals, Viterbi algorithm,  \proglang{R}}

\section{Introduction}

The package \pkg{hhsmm}, developed in the \proglang{R} language \citep{r10}, involves new tools for modeling multivariate and multi-sample time series by hidden hybrid Markov/semi-Markov models, introduced by \cite{g05}.
A hidden hybrid Markov/semi-Markov Model (HHSMM) is a model with both Markovian and semi-Markovian states. This package is available from the Comprehensive R Archive Network (CRAN) at \url{https://cran.r-project.org/package=hhsmm}.

The  hidden hybrid Markov/semi-Markov models have many applications for the situations in which there are absorbing or macro-states in the model. These flexible models decrease the time complexity of the hidden semi-Markov models, preserving their prediction power. Another important application of the  hidden hybrid Markov/semi-Markov models is in the genetics, where we aim to
analysis of the DNA sequences including long interionic zones.

Several packages are available for modeling hidden Markov and semi-Markov models. Some of the packages developed in the  \proglang{R} language are \pkg{depmixS4} \citep{vs10}, \pkg{HiddenMarkov} \citep{h06}, \pkg{msm} \citep{j07},  \pkg{hsmm} \citep{bea10} and \pkg{mhsmm} \citep{oh11}. The packages \pkg{depmixS4}, \pkg{HiddenMarkov} and \pkg{msm} only consider hidden Markov models (HMM), while the two packages  \pkg{hsmm} and \pkg{mhsmm} focus on hidden Markov and hidden semi-Markov (HSMM) models from single and multiple sequences, respectively. These packages do not include hidden hybrid Markov/semi-Markov models, which are included in the \pkg{hhsmm} package. The \pkg{mhsmm} package has some tools for fitting the HMM and HSMM models to the multiple sequences, while the \pkg{hsmm} package has not such a capability. Such a capability is preserved in the \pkg{hhsmm} package. Furthermore, the \pkg{mhsmm} package is equipped with the ability to define new emission distributions, by using the \code{mstep} functions, which is also preserved for the \pkg{hhsmm} package. In addition to all these differences, the \pkg{hhsmm} package is distinguished from \pkg{hsmm} and \pkg{mhsmm} packages in the following features:
\begin{itemize}
\item Some initialization tools are developed for initial clustering, parameter estimation, and model initialization;
\item The left-to-right models, { which are the models in which the process goes from a state to the next state and never comes back to the previous state, such as the health states of a system and the states of a phoneme in speech recognition,} are considered;
\item { The ability to initialize, fit and predict the models based on data sets containing missing values, using the EM algorithm and imputation methods, is involved; }
\item { The regime Markov/semi-Markov switching linear and additive regression model as well as the auto-regressive HHSMM  are involved; }
\item { The nonparametric estimation of the emission distribution using penalized B-splines  is added;}
\item The prediction of the future states is involved;
\item The estimation of the residual useful lifetime (RUL), { which is the remaining time to the failure of a system (the last  state of the left-to-right model, considered for the health of a system),} is developed for the left-to-right models used in the reliability theory applications;
\item The continuous sojourn time distributions are considered in their correct form;
\item The Commercial Modular Aero-Propulsion
System Simulation (\code{CMAPSS}) data set is included in the package.
\end{itemize}

There are also tools for modeling HMM in other languages. For instance, the \pkg{hmmlearn} library in \proglang{Python} or \code{hmmtrain} and \code{hmmestimate} functions in Statistics and Machine Learning Toolbox of \proglang{Matlab} are available for modeling HMM, while none of them are not suitable for modeling HSMM or HHSMM.

The remainder of the paper is organized as follows. In Section \ref{s2}, we introduce the hidden hybrid Markov/semi-Markov models (HHSMM), proposed by \cite{g05}. Section \ref{s3} presents a simple example of the HHSMM model and the \pkg{hhsmm} package. { Section \ref{s5} presents special features of the \pkg{hhsmm} package including tools for handling missing values, initialization tools, the nonparametric mixture of B-splines for estimation of the emission distribution, regime (Markov/semi-Markov) switching regression, and auto-regressive HHSMM, prediction of the future state sequence, residual useful lifetime (RUL) estimation for reliability applications, continuous-time sojourn distributions, and some other features of the hhsmm package.} Finally, the analysis of two real data sets is considered in Section \ref{rdas}, to illustrate the application of the \pkg{hhsmm} package.

\section{Hidden hybrid Markov/semi-Markov models }\label{s2}

{ Consider a sequence of observations $\{X_t\}$, which is observed for $t = 1,\ldots, T$. Assume that the distribution of $X_t$ depends on an un-observed (latent) variable $S_t$, called \emph{state}. If the sequence $\{S_t\}$ is a Markov chain of order 1, and for any $t\geq 1$, $X_t$ and $X_{t+1}$ are conditionally independent, given $S_t$, then the sequence $\{(S_t, X_t)\}$ forms a hidden Markov model (HMM). A graphical representation of the dependence structure of the HMM is shown in Figure \ref{hmmg}.

\begin{figure}
\centerline{\includegraphics[scale=0.5]{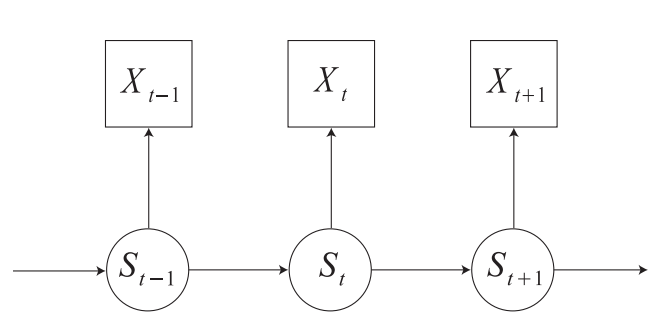}}
\caption{A graphical representation of the dependence structure of the HMM.}\label{hmmg}
\end{figure}

The parameters of an HMM are the \emph{initial probabilities} of the states, the \emph{transition probability} matrix of states, and the parameters of the conditional distribution of observations given states, which is called the \emph{emission} distribution.

The time spent in a state is called the sojourn time. In the HMM model, the sojourn time distribution is simply proved to be geometric distribution. The hidden semi-Markov model (HSMM) is similar to HMM, while the sojourn time distribution can be any other distribution, including discrete and continuous distributions with positive support, such as Poisson, negative binomial, logarithmic, nonparametric, gamma, Weibull, log-normal, etc.

}

{ The hidden hybrid Markov/semi-Markov model (HHSMM), introduced by \cite{g05}, is a combination of the HMM and HSMM models.} It is defined, for
$t=0,\ldots,\tau-1$ and $j=1,\ldots,J$,
 by the following parameters:

\begin{enumerate}
\item initial probabilities
$\pi_j = P(S_0 = j),\; \sum_{j}\pi_j = 1$,
\item  transition probabilities, which are
\begin{itemize}
\item for a semi-Markovian state $j$,
$$p_{jk} = P(S_{t+1} =k|S_{t+1} \neq j,S_t=j), \;  \forall k\neq j;\;  \sum_{k\neq j}p_{jk} = 1 ;\; p_{jj} = 0$$
\item
for a Markovian state $j$,
$$ \tilde{p}_{jk} = P(S_{t+1} = k|S_t = j);\; \sum_{k}\tilde{p}_{jk} = 1$$
\end{itemize}

By the above definition, { any \emph{absorbing state}, which is a state $j$ with $p_{jj} = 1$, is Markovian. This means that if we want to conclude an absorbing state along with some semi-Markovian states in the model, we need to use the HHSMM model. }
\item emission distribution parameters, $\theta$, for the following distribution function
$$f_j(x_t) = f(x_t | S_t = j; \; \theta)$$
\item  the sojourn time distribution is defined for semi-Markovian state $j$,
as follows
$$d_j(u) = P(S_{t+u+1} \neq j,\;S_{t+u-\nu} = j, \;\nu = 0, \ldots,u-2|S_{t+1} = j ,\; S_t \neq j), \quad  u = 1,\ldots,M_j,$$
where $M_j$ stands for an upper bound to the time spent in state $j$.
Also, the survival function of the sojourn time distribution is defined as $D_j(u) = \sum_{\nu\geq u}d_j(\nu)$.
\end{enumerate}

For a Markovian state $j$, the sojourn time distribution is the geometric distribution with the following probability mass function
$$d_j(u) = (1-\tilde{p}_{jj})\tilde{p}_{jj}^{u-1},\quad u = 1,2,\cdots $$
{ The parameter estimation of the model is performed via the \emph{EM algorithm} \citep{dea77}. The EM algorithm consists of two steps. In the first step, which is called the \emph{E-step}, the conditional expectation of the unobserved variables (states) is computed given the observations, which are called the \emph{E-step probabilities}. This step utilizes the \emph{forward-backward} algorithm to calculate the E-step probabilities. The second step is the maximization step (\emph{M-step}). In this step the parameters of the model are updated by maximizing the expectation of the logarithm of the joint probability density/mass function of the observed and unobserved data. A brief description of the EM and forward-backward algorithms, as well as the Viterbi algorithm, is given in the Appendix. The Viterbi algorithm obtains the most likely state sequence, for the HHSMM model. }

\subsection{Examples of hidden hybrid Markov/semi-Markov models}

Some examples of HHSMM models are as follows:
\begin{itemize}

\item {\bf Models with macro-states}: The macro-states are series or parallel
networks of states with common emission distribution. A semi-Markovian model can not
be used for macro-states and a hybrid Markov/semi-Markov model is a good choice in such situations \citep[see][]{cr86, dea98, g05}.

\item {\bf Models with absorbing states:}  An absorbing state is Markovian by definition. Thus, a model with an absorbing state can not be fully semi-Markovian.

\item {\bf Left to right models}: The left-to-right models are useful tools in the reliability analysis of failure systems. Another application of these models is in speech recognition, where the feature sequence extracted from a voice is modeled by a left to right model of states. The transition matrix of a left to right model is an upper triangle matrix with its final diagonal element equal to one, since the last state of a left-to-right model is absorbing. Thus, a hidden hybrid Markov/semi-Markov model might be used in such cases, instead of a hidden fully semi-Markov model.

\item {\bf Analysis of DNA sequences}: It is observed that the length of some interionic zones in
DNA sequences are approximately, geometrically distributed, while the length of other zones might deviate from the geometric distribution \citep{g05}.

\end{itemize}

\section{A simple example}\label{s3}

To illustrate the application of the \pkg{hhsmm} package for initializing, fitting, and prediction of a hidden hybrid Markov/semi-Markov model, we first propose a simple example. We emphasize that the aim of this example is not comparing different models, while this is an example to show how can we use the flexible options of the package \pkg{hhsmm} for initializing and fitting different models.

To do this, we define a model, with two Markovian and one semi-Markovian state, and 2, 3, and 2 mixture components in states 1-3, respectively, as follows. The sojourn time distribution for the semi-Markovian state is considered to be the gamma distribution (see Section \ref{cts}). The Boolean vector \code{semi} is used to define the Markovian and semi-Markovian states. Also, the mixture component proportions are defined using the parameter list \code{mix.p}.

\begin{figure}
\centerline{\includegraphics[scale=0.5]{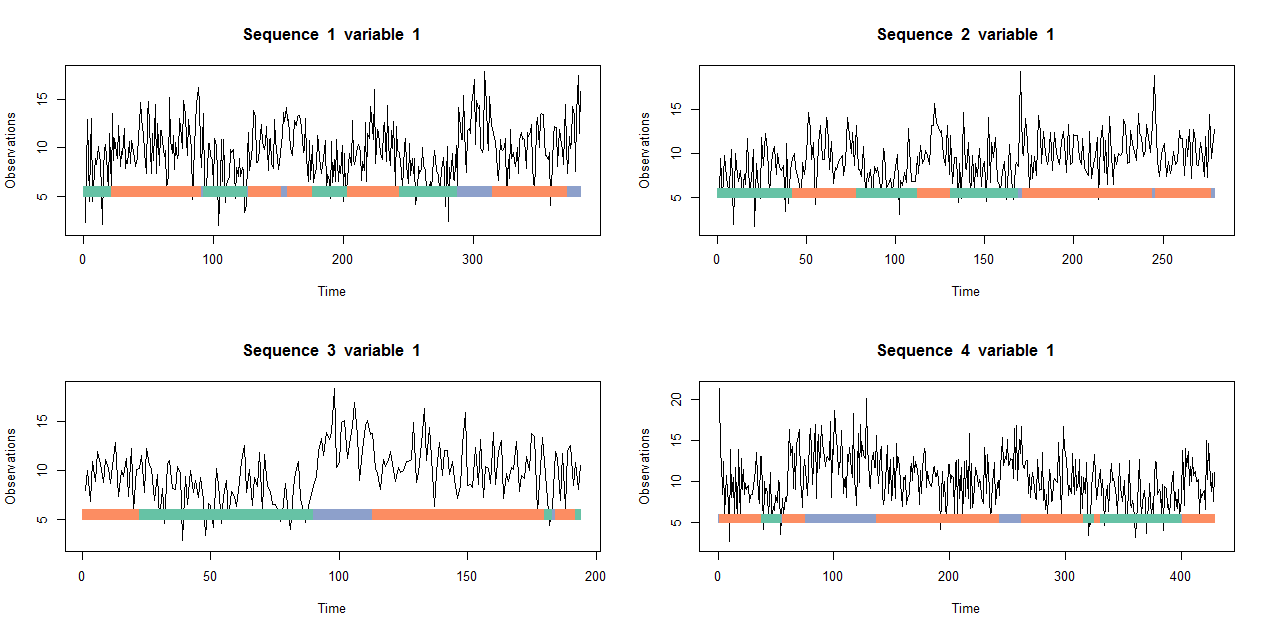}}
\caption{The plots for 4 sequences of \code{train} data set.}\label{1}
\end{figure}

\begin{figure}
\centerline{\includegraphics[scale=0.5]{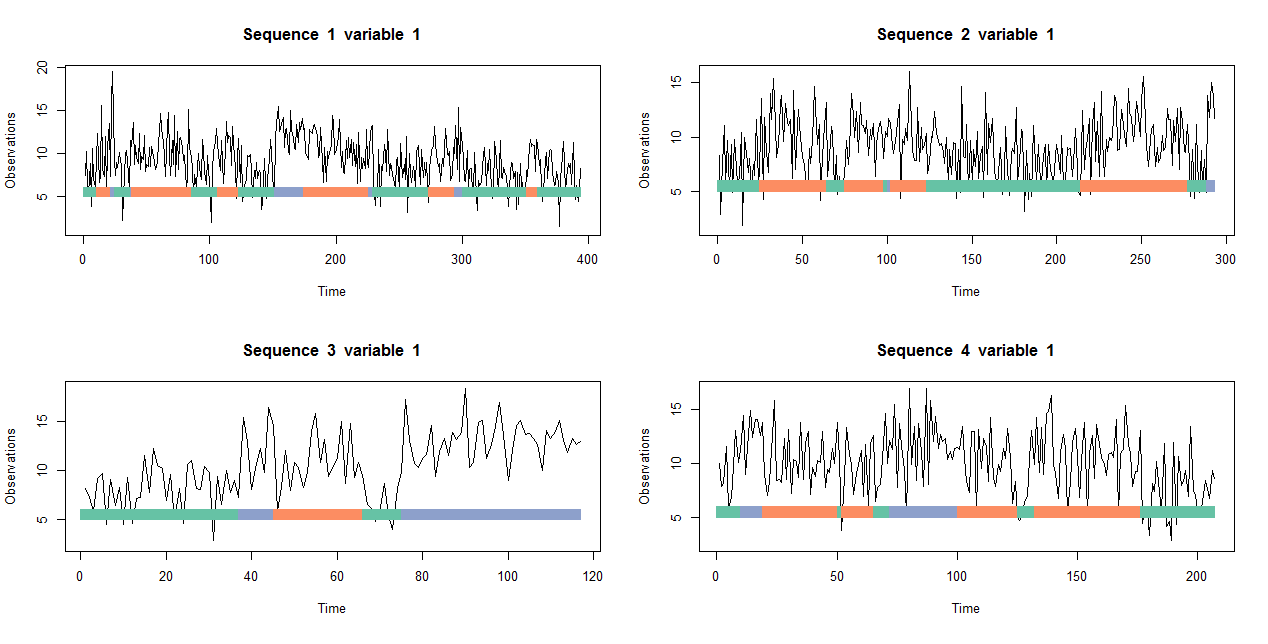}}
\caption{The plots for 4 sequences of \code{test} data set.}\label{2}
\end{figure}

\begin{lstlisting}
|\bf \color{lightgray} R$>$| J <- 3
|\bf \color{lightgray} R$>$| initial <- c(1, 0, 0)
|\bf \color{lightgray} R$>$| semi <- c(FALSE, TRUE, FALSE)
|\bf \color{lightgray} R$>$| P <- matrix(c(0.8, 0.1, 0.1, 0.5, 0, 0.5, 0.1, 0.2, 0.7),
|\bf \color{lightgray} +|     nrow = J, byrow=TRUE)
|\bf \color{lightgray} R$>$| par <- list(mu = list(list(7 , 8), list(10, 9, 11),
|\bf \color{lightgray} +|     list(12, 14)), sigma = list(list(3.8, 4.9),
|\bf \color{lightgray} +|     list(4.3, 4.2, 5.4), list(4.5, 6.1)),
|\bf \color{lightgray} +|     mix.p = list(c(0.3, 0.7), c(0.2, 0.3, 0.5), c(0.5, 0.5)))
|\bf \color{lightgray} R$>$| sojourn <- list(shape = c(0, 3, 0), scale = c(0, 10, 0),
|\bf \color{lightgray} +|     type = "gamma")
|\bf \color{lightgray} R$>$| model <- hhsmmspec(init = initial, transition = P,
|\bf \color{lightgray} +|     parms.emis = par, dens.emis = dmixmvnorm,
|\bf \color{lightgray} +|     sojourn = sojourn, semi = semi)
\end{lstlisting}
Now, we simulate \code{train} and \code{test} data sets, using the \code{simulate} function. The \code{remission} argument is considered to be \code{rmixmvnorm}, which is a function for random sample generation from mixture of multivariate normal distributions. The data sets are plotted using the \code{plot} function. The plots of the \code{train} and \code{test} data sets are presented in Figures \ref{1} and \ref{2}, respectively. Different states are distinguished with different colors in the horizontal axis.
\begin{lstlisting}
|\bf \color{lightgray} R$>$| train <- simulate(model, nsim = c(50, 40, 30, 70), seed = 1234,
|\bf \color{lightgray} +|     remission = rmixmvnorm)
|\bf \color{lightgray} R$>$| test <-  simulate(model, nsim = c(80, 45, 20, 35), seed = 1234,
|\bf \color{lightgray} +|     remission = rmixmvnorm)
|\bf \color{lightgray} R$>$| plot(train)
|\bf \color{lightgray} R$>$| plot(test)
\end{lstlisting}

In order to initialize the parameters of the HHSMM model, we first obtain an initial clustering of the \code{train} data set, using the \code{initial\_cluster} function. The \code{nstate} argument is set to 3, and the number of mixture components in the three states is set to \code{c(2,2,2)}. The \code{ltr} and \code{final.absorb} arguments is set to \code{FALSE}, which means that the model is not left-to-right and the final element of each sequence is not in an absorbing state. Thus, the \code{kmeans} algorithm \citep{ll82} is used for the initial clustering.

\begin{figure}
\centerline{\includegraphics[width=7cm]{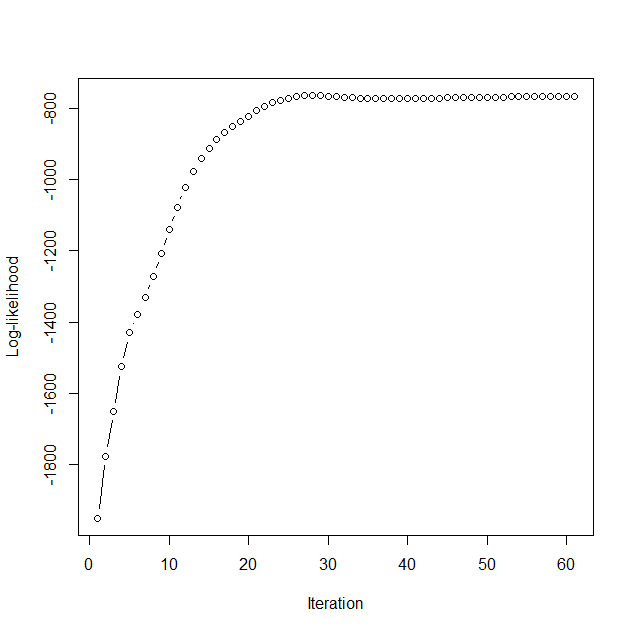}}
\caption{The log-likelihood trend during the model fitting. }\label{3}
\end{figure}

\begin{lstlisting}
|\bf \color{lightgray} R$>$| clus = initial_cluster(train, nstate = 3, nmix = c(2, 2, 2),
|\bf \color{lightgray} +|     ltr = FALSE, final.absorb = FALSE, verbose = FALSE)
\end{lstlisting}
Now, we initialize the model parameters using the \code{initialize\_model} function. The initial clustering output \code{clus} is used for estimation of the parameters. The \code{sojourn} time distribution is set to \code{"gamma"} distribution. First, we use the true value of the \code{semi} vector for modeling. Thus, the initialized model is a hidden hybrid Markov/semi-Markov model.
\begin{lstlisting}
|\bf \color{lightgray} R$>$| semi <- c(FALSE, TRUE, FALSE)
|\bf \color{lightgray} R$>$| initmodel1 = initialize_model(clus = clus, sojourn = "gamma",
|\bf \color{lightgray} +|     M = max(train$N), semi = semi)
\end{lstlisting}
The model is then fitted using the \code{hhsmmfit} function as follows. The initialized model \code{initmodel1} is used as the start value.
\begin{lstlisting}
|\bf \color{lightgray} R$>$| fit1 = hhsmmfit(x = train, model = initmodel1, M = max(train$N),
|\bf \color{lightgray} +|     par = list(verbose = FALSE))
\end{lstlisting}
The log-likelihood trend can also be extracted and plotted as follows. This plot is presented in Figure \ref{3}.
\begin{lstlisting}
|\bf \color{lightgray} R$>$| plot(fit1$loglik[-1], type = "b", ylab = "Log-likelihood",
|\bf \color{lightgray} +|     xlab = "Iteration")
\end{lstlisting}
{One can observe, for instance, the estimated initial probabilities, transition matrix, and the estimated parameters of the sojourn distribution as follows.
\begin{lstlisting}
|\bf \color{lightgray} R$>$| fit1$model$init
[1] 0.6261528 0.1144543 0.2593930
|\bf \color{lightgray} R$>$| fit1$model$transition
             [,1]       [,2]       [,3]
[1,] 9.638477e-01 0.02294345 0.01320884
[2,] 4.571514e-01 0.00000000 0.54284857
[3,] 3.460872e-10 0.09880954 0.90119046
|\bf \color{lightgray} R$>$| fit1$model$sojourn
$shape
[1] 0.0000000 0.9714843 0.0000000
$scale
[1]  0.0000 21.2017  0.0000
$type
[1] "gamma"
 \end{lstlisting}}
The state sequence is now predicted using the default method \code{"viterbi"} of the \code{predict} function for the \code{test} data set. Because of the displacement property of the states, the homogeneity of the predicted states is computed using the \code{homogeneity} function for three states. { Since the states are indeed clusters, the homogeneity measures, which are used for clustering, are useful for measuring the homogeneity of two sequences of state. The homogeneity of a specified cluster (state) in two sequences, is defined as the percentage of members of both sequences that are in the same cluster (state) in both sequences.} The output of the \code{homogeneity} function shows the {homogeneity} percent of two sequences of states.
\begin{lstlisting}
|\bf \color{lightgray} R$>$| yhat1 <- predict(fit1, test)
|\bf \color{lightgray} R$>$| homogeneity(yhat1$s , test$s)
[1] 0.9191686 0.8564920 0.7553957
\end{lstlisting}
Now, we initialize and fit a fully Markovian model (HMM) by setting \code{semi} to \code{c(FALSE,FALSE,FALSE)}. The same clustering output \code{clus} can be used here.
\begin{lstlisting}
|\bf \color{lightgray} R$>$| semi <- c(FALSE, FALSE, FALSE)
|\bf \color{lightgray} R$>$| initmodel2 = initialize_model(clus = clus, M = max(train$N),
|\bf \color{lightgray} +|     semi = semi)
\end{lstlisting}
The model is again fitted using \code{hhsmmfit} function.
\begin{lstlisting}
|\bf \color{lightgray} R$>$| fit2 = hhsmmfit(x = train, model = initmodel2, M = max(train$N),
|\bf \color{lightgray} +|     par = list(lock.init = TRUE, verbose = FALSE))
\end{lstlisting}
We can compare some of the estimated parameters of this model with those of the previous one.
\begin{lstlisting}
|\bf \color{lightgray} R$>$| fit2$model$init
[1] 0.3333333 0.3333333 0.3333333
|\bf \color{lightgray} R$>$| fit2$model$transition
             [,1]       [,2]       [,3]
[1,] 9.681322e-01 0.01670787 0.01515991
[2,] 1.695894e-02 0.96554352 0.01749754
[3,] 4.716631e-16 0.08631537 0.91368463
\end{lstlisting}
Now, we predict the state sequence of the fitted model and compute its homogeneity with the true state sequence.
\begin{lstlisting}
|\bf \color{lightgray} R$>$| yhat2 <- predict(fit2, test)
|\bf \color{lightgray} R$>$| homogeneity(yhat2$s , test$s)
[1] 0.9237875 0.8609272 0.8400000
 \end{lstlisting}
Finally, we initialize and fit a full semi-Markov model (HSMM) to the \code{train} data set, by setting \code{semi} to \code{c(TRUE,TRUE,TRUE)}. The \code{"gamma"} distribution is considered as the sojourn time distribution for all states.
\begin{lstlisting}
|\bf \color{lightgray} R$>$| semi <- c(TRUE, TRUE, TRUE)
|\bf \color{lightgray} R$>$| initmodel3 = initialize_model(clus=clus, sojourn = "gamma",
|\bf \color{lightgray} +|     M = max(train$N), semi = semi)
|\bf \color{lightgray} R$>$| fit3 = hhsmmfit(x = train, model = initmodel3, M = max(train$N),
|\bf \color{lightgray} +|     par = list(verbose = FALSE))
|\bf \color{lightgray} R$>$| fit3$model$transition
             [,1]       [,2]      [,3]
[1,] 0.000000e+00 0.02704853 0.9729515
[2,] 2.242357e-01 0.00000000 0.7757643
[3,] 4.037597e-06 0.99999596 0.0000000
|\bf \color{lightgray} R$>$| fit3$model$sojourn
$shape
[1] 4.2375333 3.4556658 0.1567049
$scale
[1]  8.802259  4.408482 31.021809
$type
[1] "gamma"
\end{lstlisting}
The prediction and homogeneity computation for this model is done as follows.
\begin{lstlisting}
|\bf \color{lightgray} R$>$| yhat3 <- predict(fit3, test)
|\bf \color{lightgray} R$>$| homogeneity(yhat3$s , test$s)
[1] 0.9232737 0.8069414 0.7358491
\end{lstlisting}

\section{Special features of the package}\label{s5}

The \pkg{hhsmm} package has several special features, which are described in the following subsections.

{\subsection{Handling missing values}

The \pkg{hhsmm} package is equipped with tools for handling data sets with missing values. A special imputation algorithm is used in the \code{initial\_cluster} function. This algorithm, imputes a completely missed row of the data with the average of its previous and next rows, while if some columns are missed, the predictive mean matching method of the function \code{mice} from package \pkg{mice} \citep{mice}, with $m=1$, is used to initially impute the missing values. After performing the initial clustering and initial estimation of the parameters of the model, the \code{miss\_mixmvnorm\_mstep} function is considered, as the M-step function of the EM algorithm, for initializing and fitting the model. The function \code{miss\_mixmvnorm\_mstep} includes computation of the conditional means and conditional second moments of the missing values given observed values in each iteration of the EM algorithm and updating the parameters of the Gaussian mixture emission distribution, using the \code{cov.miss.mix.wt} function. Furthermore, an approximation of the mixture component weights using the observed values and conditional means of the missing values given observed values is used in each iteration. The values of the emission density function, used in the E-step of the EM algorithm are computed by replacing the missing values with their conditional means given the observed values.

Here, we provide a simple example to examine the performance of the aforementioned method. First, we define a model with three states and two variables.
\begin{lstlisting}
|\bf \color{lightgray} R$>$| J <- 3
|\bf \color{lightgray} R$>$| initial <- c(1, 0, 0)
|\bf \color{lightgray} R$>$|  semi <- c(FALSE, TRUE, FALSE)
|\bf \color{lightgray} R$>$|  P <- matrix(c(0.8, 0.1, 0.1, 0.5, 0, 0.5, 0.1, 0.2, 0.7),
|\bf \color{lightgray} +|     nrow = J, byrow = TRUE)
|\bf \color{lightgray} R$>$|  par <- list(mu = list(list(c(7, 17), c(8, 18)),
|\bf \color{lightgray} +|     list(c(15, 25), c(14, 24), c(16, 16)),
|\bf \color{lightgray} +|     list(c(0, 10), c(2, 12))),
|\bf \color{lightgray} +|     sigma = list(list(diag(c(2.8, 4.8)), diag(c(3.9, 5.9))),
|\bf \color{lightgray} +|     list(diag(c(3.3, 5.3)), diag(c(3.2, 5.2)),
|\bf \color{lightgray} +|     diag(c(4.4, 6.4))), list(diag(c(3.5, 5.5)), diag(c(5.1, 7.1)))),
|\bf \color{lightgray} +|     mix.p = list(c(0.3, 0.7), c(0.2, 0.3, 0.5), c(0.5, 0.5)))
|\bf \color{lightgray} R$>$| sojourn <- list(shape = c(0, 3, 0), scale = c(0, 10, 0),
|\bf \color{lightgray} +|     type = "gamma")
|\bf \color{lightgray} R$>$| model <- hhsmmspec(init = initial, transition = P,
|\bf \color{lightgray} +|     parms.emis = par, dens.emis = dmixmvnorm,
|\bf \color{lightgray} +|     sojourn = sojourn, semi = semi)
\end{lstlisting}
Now, we simulate the complete train and test data sets.
\begin{lstlisting}
|\bf \color{lightgray} R$>$| train <- simulate(model, nsim = c(10, 8, 8, 18), seed = 1234,
|\bf \color{lightgray} +|        remission = rmixmvnorm)
|\bf \color{lightgray} R$>$|  test <- simulate(model, nsim = c(8, 6, 6, 15), seed = 1234,
|\bf \color{lightgray} +|        remission = rmixmvnorm)
\end{lstlisting}
First, we initialize and fit the model with complete data sets. To do this, we first use the \code{initial\_cluster} to provide an initial clustering of the \code{train} data set.
\begin{lstlisting}
|\bf \color{lightgray} R$>$| clus = initial_cluster(train, nstate = 3, nmix = c(2, 2, 2),
|\bf \color{lightgray} +|        ltr = FALSE, final.absorb = FALSE, verbose = FALSE)
\end{lstlisting}
Now, we initialize and fit the model.
\begin{lstlisting}
|\bf \color{lightgray} R$>$|  semi <- c(FALSE, TRUE, FALSE)
|\bf \color{lightgray} R$>$|  initmodel1 = initialize_model(clus = clus, sojourn = "gamma",
|\bf \color{lightgray} +|       M = max(train$N), semi = semi)
|\bf \color{lightgray} R$>$| fit1 = hhsmmfit(x = train, model = initmodel1, M = max(train$N),
|\bf \color{lightgray} +|       par = list(verbose = FALSE))
\end{lstlisting}
Finally, we predict the state sequence of the test data set, using the \code{predict.hhsmm} function and the default \code{"viterbi"} method.
\begin{lstlisting}
|\bf \color{lightgray} R$>$| yhat1 <- predict(fit1, test)
\end{lstlisting}
To examine the tools for modeling the data sets with missing values, we randomly select some elements of the \code{train} and \code{test} data sets and replace them with \code{NA}, as follows.
\begin{lstlisting}
|\bf \color{lightgray} R$>$| p = ncol(train$x)
|\bf \color{lightgray} R$>$| n = nrow(train$x)
|\bf \color{lightgray} R$>$| sammissless = sample(1:n, trunc(n / 10))
|\bf \color{lightgray} R$>$| sammissall = sample(1:n, trunc(n / 20))
|\bf \color{lightgray} R$>$| misrat = matrix(rbinom(trunc(n / 10) * p, 1, 0.2),
|\bf \color{lightgray} +|     trunc(n / 10), p)
|\bf \color{lightgray} R$>$|  train$x[sammissall, ] <- NA
|\bf \color{lightgray} R$>$|  for(i in 1:trunc(n / 10))
|\bf \color{lightgray} +|     train$x[sammissless[i], misrat[i,] == 1] <- NA
|\bf \color{lightgray} R$>$|  nt = nrow(test$x)
|\bf \color{lightgray} R$>$|  sammissless = sample(1:nt, trunc(nt / 12))
|\bf \color{lightgray} R$>$| sammissall = sample(1:nt, trunc(nt / 25))
|\bf \color{lightgray} R$>$|  misrat = matrix(rbinom(trunc(nt / 12)*p, 1, 0.15),
|\bf \color{lightgray} +|     trunc(nt / 12), p)
|\bf \color{lightgray} R$>$|  test$x[sammissall,] <- NA
|\bf \color{lightgray} R$>$|  for(i in 1:trunc(nt/12))
|\bf \color{lightgray} +|     test$x[sammissless[i], misrat[i, ] == 1] <- NA
\end{lstlisting}
Now, we provide the initial clustering of the incomplete \code{train} data set using the \code{initial\_cluster} function.
\begin{lstlisting}
|\bf \color{lightgray} R$>$| clus = initial_cluster(train, nstate = 3, nmix = c(2, 2, 2),
|\bf \color{lightgray} +|     ltr = FALSE, final.absorb = FALSE, verbose = FALSE)
\end{lstlisting}
We can observe that the output of the \code{initial\_cluster} function contains a flag that indicates the missingness in the data set.
\begin{lstlisting}
|\bf \color{lightgray} R$>$| clus$miss
TRUE
\end{lstlisting}
Now, we initialize and fit the model for the incomplete data set.
\begin{lstlisting}
|\bf \color{lightgray} R$>$|  semi <- c(FALSE, TRUE, FALSE)
|\bf \color{lightgray} R$>$|  initmodel2 = initialize_model(clus = clus, sojourn = "gamma",
|\bf \color{lightgray} +|     M = max(train$N), semi = semi)
|\bf \color{lightgray} R$>$|  fit2 = hhsmmfit(x = train, model = initmodel2,
|\bf \color{lightgray} +|     M = max(train$N), par = list(lock.init = TRUE, verbose = FALSE))
\end{lstlisting}
similarly, we predict the state sequence of the incomplete test data set, using the \code{predict.hhsmm} function.
\begin{lstlisting}
|\bf \color{lightgray} R$>$| yhat2 <- predict(fit2, test)
\end{lstlisting}
We can observe that the homogeneity of the predictions of the complete and incomplete data sets are very close to each other.
\begin{lstlisting}
|\bf \color{lightgray} R$>$| homogeneity(yhat1$s, test$s)
[1] 0.8487395 0.9793814 0.0000000
|\bf \color{lightgray} R$>$| homogeneity(yhat2$s, test$s)
[1] 0.9830508 0.8595041 0.0000000
\end{lstlisting}}

\subsection{Tools and methods for initializing model}\label{initsec}

To initialize the HHSMM model, we need to obtain an initial clustering for the train data set. For a  left to right  model (option \code{ltr = TRUE} of the \code{initial\_cluster} function), we propose Algorithm \ref{alg2}, which uses Algorithm \ref{alg1}, for a  left to right  initial clustering, which are included in the function \code{ltr\_clus} of the \pkg{hhsmm} package. These algorithms use Hotelling's T-squared test statistic as the distance measure for clustering. The simulations and real data analysis show that the starting values obtained by the proposed algorithm perform well for a left to right model (see Section \ref{rdas} for a real data application). If the model is not a left to right model, then the usual K-means algorithm is used for clustering. Furthermore, the K-means algorithm is used within each initial state to cluster data for mixture components. The number of mixture components can be determined automatically, using the option \code{nmix = "auto"}, by analysis of the within sum of squares obtained from the \code{kmeans} function. The number of starting values of the \code{kmeans} is set to 10, for the stability of the results. The initial clustering is performed using the \code{initial\_cluster} function.

\begin{algorithm}
    \caption{The left to right clustering algorithm for two clusters.}
    \label{alg1}
    \begin{algorithmic}
\item For $s = 2,\ldots,k-2$ consider partitions $\{ 1, \ldots ,k\} = \{ 1, \ldots, s\} \cup \{ s+1, \ldots, k\} $ and compute the means
$$\bar{X}_{1s} = \frac{1}{s} \sum_{i=1}^s X_i, \quad \bar{X}_{2s} = \frac{1}{k-s} \sum_{i=s+1}^k X_i,$$
the variance-covariance matrices
$$\Sigma_{1s} = \frac{1}{s-1} \sum_{i=1}^s (X_i-\bar{X}_{1s})(X_i-\bar{X}_{1s})^\top ,
\Sigma_{2s} = \frac{1}{k-s-1} \sum_{i=s+1}^k (X_i-\bar{X}_{2s})(X_i-\bar{X}_{2s})^\top,$$
and the standardized distances (Hotelling's T-squared test statistic)
$$d_s = \frac{(s(k-s)/k)(k-p-1)}{(k-2)p}(\bar{X}_{1s}-\bar{X}_{2s})^\top\Sigma_{ps}^{-1} (\bar{X}_{1s}-\bar{X}_{2s}),$$
where $$\Sigma_{ps}=\frac{(s-1)\Sigma_{1s}+(k-s-1)\Sigma_{2s}}{k-2}.$$
\item Let $s^* = \arg\max_s d_s$.
\item If $d_{s^*} > F_{(0.05;p,k-1-p)}$, the clusters would be $\{ 1, \ldots, s^*\} $ and $ \{ s^*+1, \ldots, k\}$, otherwise, no clustering will be done, where $F_{(0.05;p,k-1-p)}$
stands for the 95th percentile of the F distribution with $p$
 and $k-1-p$ degrees of freedom.
    \end{algorithmic}
\end{algorithm}

After obtaining the initial clustering, the initial estimates of the parameters of the mixture of multivariate normal emission distribution are obtained. Furthermore, the parameters of the sojourn time distribution is obtained by running the method of moments estimation algorithms on the time duration observations of the initial clustering of each state. If we set \code{sojourn = "auto"} in the \code{initialize\_model} function, the best sojourn time distribution is selected from the list of available sojourn time distributions, using the Chi-square goodness of fit testing on the initial cluster data of all states.

\begin{algorithm}
    \caption{The left to right clustering algorithm for $K>2$ clusters.}
    \label{alg2}
    \begin{algorithmic}
\item Let Nclust = 1. While Nclust $< K$ and the clusters change, do
    \begin{itemize}
\item for all clusters run Algorithm \ref{alg1} to obtain two clusters
    \end{itemize}
\item If Nclust $> K$, while Nclust $= K$ do
    \begin{itemize}
\item merge clusters with minimum $d_{s^*}$ values its closest neigbour on its right or left.
    \end{itemize}
    \end{algorithmic}
\end{algorithm}

{
\subsection{Nonparametric mixture of B-splines emission}

Usually, the emission distribution belongs to a parametric family of distributions. Although the mixture of multivariate normals is shown to be a good choice in many practical situations, there are also examples in which this class of emission distribution fails to model the skewness and tail weight of the data set. Furthermore, the choice of the number of components of the mixture distribution in each state is a challenge of using mixture of multivariate normals as the emission distribution. As an alternative to parametric emission distribution, HMMs and HSMMs with non-parametric estimates of state-dependent distributions are shown to be more parsimonious in terms of the number of states, easier to interpret, and well fitted to the data \citep{let15,aea19}. The proposed nonparametric estimation approach of \cite{let15} and \cite{aea19} is based on the idea of representing the densities of the emission distributions as linear combinations of B-spline basis functions, by adding a smoothing penalty term to the quasi-log-likelihood function. In this model, the emission distribution is defined as follows
\begin{equation}\label{nmbse}
f_j(x) = \sum_{k=-K}^{K} a_{j,k} \phi_k(x), \quad j=1,\ldots,J,
\end{equation}
where $\{\phi_{-K}(\cdot),\ldots,\phi_{K}(\cdot)\}$ is a sequences of B-splines and $ \{a_{j,k}\}$ is the sequences of unknown coefficients to be estimated. These parameters are estimated in the M-step of the EM algorithm, by maximizing the following penalized quasi-log-likelihood function
\begin{equation}\label{pqllf}
\ell_P^{\rm HHSMM}(\theta,\lambda) = log(L^{\rm HHSMM}(\theta)) - \frac{1}{2}\sum_{j=1}^J \lambda_j \sum_{k=-K+2}^{K} (\Delta^2 a_{j,k})^2,
\end{equation}
in which $L^{HHSMM}(\theta)$ the quasi-likelihood of the HHSMM model, $\theta$ is the parameters of the model,
$\Delta a_k = a_k - a_{k-1}$, $\Delta^2 a_k = \Delta (\Delta a_k)$, and
 $\lambda_1,\ldots,\lambda_J$ are the smoothing parameters, which are estimated as follows \citep{sk12}
$$
\hat{\lambda}_j = \frac{{\rm df}(\hat{\lambda}_j) - p}{\sum_{k=-K+2}^{K}(\Delta^2 \hat{a}_{j,k})^2},
$$
where $p$ is the dimension of the data,
$${\rm df}(\hat{\lambda}_j)={\rm tr}\left(H^{-1}(\hat{a}_j;\lambda_j = \hat{\lambda}_j) H(\hat{a}_j;\lambda_j = 0)\right),$$
and $H(\hat{a};\lambda)$ is the hessian matrix of the log-quasi-likelihood at $\hat{a}$ with the specified value of $\lambda$.

To illustrate the application of the \pkg{hhsmm} package with a nonparametric mixture of B-splines emission distribution, we present a simple simulated data example. To do this, we first simulate data from an HHSMM model with a mixture of multivariate normals as the emission distribution, as follows.

\begin{lstlisting}
|\bf \color{lightgray} R$>$|  J <- 3
|\bf \color{lightgray} R$>$| initial <- c(1,0,0)
|\bf \color{lightgray} R$>$|  semi <- c(FALSE,TRUE,FALSE)
|\bf \color{lightgray} R$>$|  P <- matrix(c(0.8, 0.1, 0.1, 0.5, 0, 0.5, 0.1, 0.2, 0.7),
|\bf \color{lightgray} +|     nrow = J, byrow = TRUE)
|\bf \color{lightgray} R$>$| par <- list(mu = list(
|\bf \color{lightgray} +|     list(c(7, 17), c(8, 18)),
|\bf \color{lightgray} +|     list(c(15, 25), c(14, 24), c(16, 16)),
|\bf \color{lightgray} +|     list(c(0, 10), c(2, 12))),
|\bf \color{lightgray} +|     sigma = list(list(diag(c(2.8, 4.8)), diag(c(3.9, 5.9))),
|\bf \color{lightgray} +|     list(diag(c(3.3, 5.3)), diag(c(3.2, 5.2)), diag(c(4.4, 6.4))),
|\bf \color{lightgray} +|     list(diag(c(3.5, 5.5)), diag(c(5.1, 7.1)))),
|\bf \color{lightgray} +|     mix.p = list(c(0.3, 0.7),
|\bf \color{lightgray} +|     c(0.2, 0.3, 0.5),c(0.5, 0.5)))
|\bf \color{lightgray} R$>$|  sojourn <- list(shape = c(0,3,0), scale = c(0,10,0),
|\bf \color{lightgray} +|     type = "gamma")
|\bf \color{lightgray} R$>$|  model <- hhsmmspec(init = initial, transition = P,
|\bf \color{lightgray} +|     parms.emis = par, dens.emis = dmixmvnorm,
|\bf \color{lightgray} +|     sojourn = sojourn, semi = semi)
|\bf \color{lightgray} R$>$|  train <- simulate(model, nsim = c(10, 8, 8, 18), seed = 1234,
|\bf \color{lightgray} +|     remission = rmixmvnorm)
|\bf \color{lightgray} R$>$|  test <- simulate(model, nsim = c(8, 6, 6, 15), seed = 1234,
|\bf \color{lightgray} +|      remission = rmixmvnorm)
\end{lstlisting}

Now, we obtain an initial clustering of the data set using the \code{initial\_cluster} function. Note that for a nonparametric
emission distribution, we have no mixture components and we should use the option \code{nmix = NULL}.

\begin{lstlisting}
|\bf \color{lightgray} R$>$|  clus = initial_cluster(train, nstate = 3, nmix = NULL,
|\bf \color{lightgray} +|     ltr = FALSE, final.absorb = FALSE, verbose = FALSE)
\end{lstlisting}

In order to initialize a HHSMM with non-parametric estimates of the emission distribution, we use the \code{initialize\_model} function with the arguments \code{mstep = nonpar\_mstep} and \code{dens.emission = dnonpar}, as follows.

\begin{lstlisting}
|\bf \color{lightgray} R$>$|  semi <- c(FALSE, TRUE, FALSE)
|\bf \color{lightgray} R$>$|  initmodel1 = initialize_model(clus = clus, mstep = nonpar_mstep,
|\bf \color{lightgray} +|     dens.emission = dnonpar, sojourn = "gamma", M = max(train$N),
|\bf \color{lightgray} +|     semi = semi)
\end{lstlisting}

Now, we can use the \code{hhsmmfit} function to fit the model.

\begin{lstlisting}
|\bf \color{lightgray} R$>$|  fit1 = hhsmmfit(x = train, model = initmodel1, M = max(train$N),
|\bf \color{lightgray} +|     par = list(verbose = FALSE))
\end{lstlisting}

Finally, we predict the state sequence of the test data and compute the homogeneity of the predicted sequence and the reals sequence as follows.

\begin{lstlisting}
|\bf \color{lightgray} R$>$| yhat1 <- predict(fit1, test)
|\bf \color{lightgray} R$>$| homogeneity(yhat1$s, test$s)
[1] 0.9210526 0.8508772 0.8750000
\end{lstlisting}

As one can see from the output of the \code{homogeneity} function, the fitted model has a high precision for the prediction of the state sequence of the new data set.

{
\subsection{Regime (Markov/semi-Markov) switching regression model}

\cite{kea08} considered the following Gaussian regime-switching model
\begin{equation}\label{rsrm}
y_{t} = x_{t}^T \beta_{s_t} + \sigma_{s_t}\epsilon_t,
\end{equation}
where $y_{t}$ is the response variable, $x_{t}$ is a vector of covariates,
which may include lagged values of $y_{t}$ (auto-regressive HHSMM, see the next subsection), $s_t$ is the state, and $\epsilon_t$ is the regression error, which is assumed to be distributed as standard normal distribution, for $t=1,\ldots,T$. Model \eqref{rsrm} can easily be extended to the case of multivariate responses and also to the case of mixture of multivariate normals. The difference between the regime-switching model \eqref{rsrm} and the HHSMM model is that, instead of using the density of $y_{t}$ given $s_t$ in the likelihood function, we use the conditional density of $y_{t}$ given $x_{t}$ and $s_t$. A graphical representation of the regime-switching model is presented in Figure \ref{msregg}. The parameters of the regime switching regression
model can be estimated using the EM algorithm. 

\begin{figure}
\centerline{\includegraphics[scale=0.35]{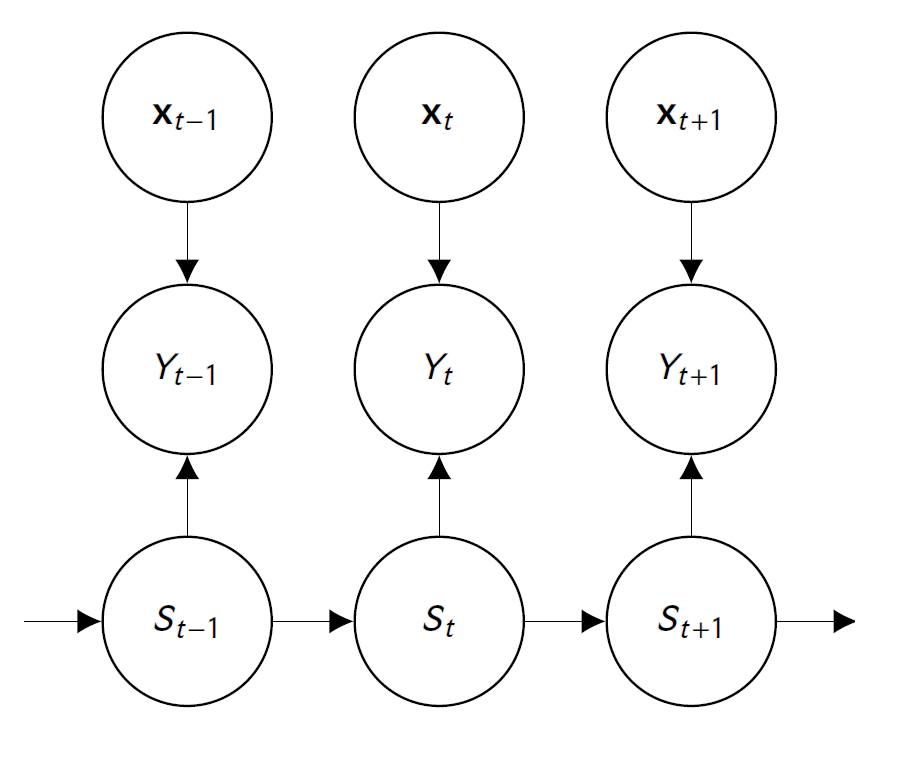}}
\caption{Graphical representation of the regime-switching model.}\label{msregg}
\end{figure}

\cite{let18} considered an extension of the model \eqref{rsrm} to the following additive regime-switching model
\begin{equation}\label{rsrm2}
y_{t} = \mu_{s_t} + \sum_{j=1}^p f_{j,s_t}(x_{j,t}) + \sigma_{s_t}\epsilon_t,
\end{equation}
where $f_{j,s_t}(\cdot),\; j=1,...,p$ are unknown regression functions for $p$ covariates. They utilized the penalized B-splines for estimation of the regression functions.

The estimation of extensions of models \eqref{rsrm} and \eqref{rsrm2} is considered in \pkg{hhsmm} package, using the \code{mixlm\_mstep} and \code{additive\_reg\_mstep} functions, respectively, as the M-step estimation and \code{dmixlm} and \code{dnorm\_additive\_reg} functions, respectively, which define mixture of multivariate normals and multivariate normal densities, respectively, as the conditional density of the responses given the covariates. The response variables are determined using the argument \code{resp.ind} in all of these functions, with its default value equal to one, which means that the first column of the input matrix \code{x}, is the univariate  response variable.

To illustrate usage of these functions in the \pkg{hhsmm} package, we present the following simple simulated data example. First, we simulate data using the function \code{simulate.hhsmmspec}, using the argument \code{remission = rmixlm}, \code{covar = list(mean = 0, cov = 1)}. The argument \code{covar} is indeed an argument of the \code{rmixlm} function, which is either a function which generates the covariate vector or a list containing the mean vector and the variance-covariance matrix of covariates to be generated from multivariate normal distribution. The \code{rmixlm} is a function for generation of the data from mixture of linear models, for each state. The list of parameters of this emission distribution consist of the following items:
\begin{itemize}
\item \code{intercept}, a list of the intercepts of the regression models for each state and each mixture component,
\item \code{coefficient}, a list of the coefficient vectors/matrices of the regression models for each state and each mixture component,
\item \code{csigma}, a list of the conditional variances/variance-covariance matrices of the response for each state and each mixture component,
\item \code{mix.p}, a list of mixture component probabilities for each state.
\end{itemize}

First, we define the model parameters and simulate the data as follows.

\begin{lstlisting}
|\bf \color{lightgray} R$>$| J <- 3
|\bf \color{lightgray} R$>$| initial <- c(1, 0, 0)
|\bf \color{lightgray} R$>$| semi <- rep(FALSE, 3)
|\bf \color{lightgray} R$>$| P <- matrix(c(0.5, 0.2, 0.3, 0.2, 0.5, 0.3, 0.1, 0.4, 0.5),
|\bf \color{lightgray} +|     nrow = J, byrow = TRUE)
|\bf \color{lightgray} R$>$| par <- list(intercept = list(3, list(-10, -1), 14),
|\bf \color{lightgray} +|     coefficient = list(-1, list(1, 5), -7),
|\bf \color{lightgray} +|     csigma = list(1.2, list(2.3, 3.4), 1.1),
|\bf \color{lightgray} +|     mix.p = list(1, c(0.4, 0.6), 1))
|\bf \color{lightgray} R$>$| model <- hhsmmspec(init = initial, transition = P,
|\bf \color{lightgray} +|     parms.emis = par, dens.emis = dmixlm, semi = semi)
|\bf \color{lightgray} R$>$| train <- simulate(model, nsim = c(20, 30, 42, 50), seed = 1234,
|\bf \color{lightgray} +|     remission = rmixlm, covar = list(mean = 0, cov = 1))
\end{lstlisting}

Now, we obtain an initial clustering of the data using \code{initial\_cluster} function, with the argument \code{regress = TRUE}, which is essential for estimation the parameters of the regime switching regression models. By letting \code{regress = TRUE} and \code{ltr = FALSE} the \code{initial\_cluster} function uses an algorithm similar to that of
\cite{ll82} for the K-means method, by fitting linear regression models instead of computing simple means, in each iteration of an algorithm. When using \code{regress = TRUE} and \code{ltr = TRUE}, an algorithm similar to that, described in section \ref{initsec} is used for left-to-right clustering, by using regression coefficients instead of mean vectors, and the associated Hotteling's T-square statistic.

\begin{lstlisting}
|\bf \color{lightgray} R$>$| clus = initial_cluster(train = train, nstate = 3,
|\bf \color{lightgray} +|     nmix = 2, ltr = FALSE, final.absorb = FALSE,
|\bf \color{lightgray} +|     verbose = FALSE, regress = TRUE)
\end{lstlisting}

We initialize the model, using the \code{initialize\_model} function, with arguments \code{mstep = mixlm\_mstep}, which is a function for M-step estimation of the EM algorithm in the regime switching regression model, and \code{dens.emission = dmixlm}, which is a function for computation of the probability density function of the mixture Gaussian linear model,
for a specified observation vector, a specified state and a specified model's parameters. Next, we fit the model, by using the \code{hhsmmfit} function.

\begin{lstlisting}
|\bf \color{lightgray} R$>$| initmodel = initialize_model(clus = clus, mstep = mixlm_mstep,
|\bf \color{lightgray} +|     dens.emission = dmixlm, sojourn = NULL, semi = rep(FALSE, 3),
|\bf \color{lightgray} +|     M = max(train$N), verbose = FALSE)
|\bf \color{lightgray} R$>$| fit1 = hhsmmfit(x = train, model = initmodel, mstep = mixlm_mstep,
|\bf \color{lightgray} +|     M = max(train$N), par = list(lock.init = TRUE, verbose = FALSE))
\end{lstlisting}

The plots of the clustered data as well as the estimated regime-switching regression model lines are then plotted as follows. The resulting plot is shown in Figure \ref{regress}.

\begin{lstlisting}
|\bf \color{lightgray} R$>$| plot(train$x[,1] ~ train$x[,2], col = train$s, pch = 16,
|\bf \color{lightgray} +|     xlab = "x", ylab = "y")
|\bf \color{lightgray} R$>$| abline(fit1$model$parms.emission$intercept[[1]][[1]],
|\bf \color{lightgray} +|     fit1$model$parms.emission$coefficient[[1]][[1]], col = 1)
|\bf \color{lightgray} R$>$| abline(fit1$model$parms.emission$intercept[[1]][[2]],
|\bf \color{lightgray} +|     fit1$model$parms.emission$coefficient[[1]][[2]], col = 1)
|\bf \color{lightgray} R$>$| abline(fit1$model$parms.emission$intercept[[2]][[1]],
|\bf \color{lightgray} +|     fit1$model$parms.emission$coefficient[[2]][[1]], col = 2)
|\bf \color{lightgray} R$>$| abline(fit1$model$parms.emission$intercept[[2]][[2]],
|\bf \color{lightgray} +|     fit1$model$parms.emission$coefficient[[2]][[2]], col = 2)
|\bf \color{lightgray} R$>$| abline(fit1$model$parms.emission$intercept[[3]][[1]],
|\bf \color{lightgray} +|     fit1$model$parms.emission$coefficient[[3]][[1]], col = 3)
|\bf \color{lightgray} R$>$| abline(fit1$model$parms.emission$intercept[[3]][[2]],
|\bf \color{lightgray} +|     fit1$model$parms.emission$coefficient[[3]][[2]], col = 3)
\end{lstlisting}

\begin{figure}
\centerline{\includegraphics[scale=0.6]{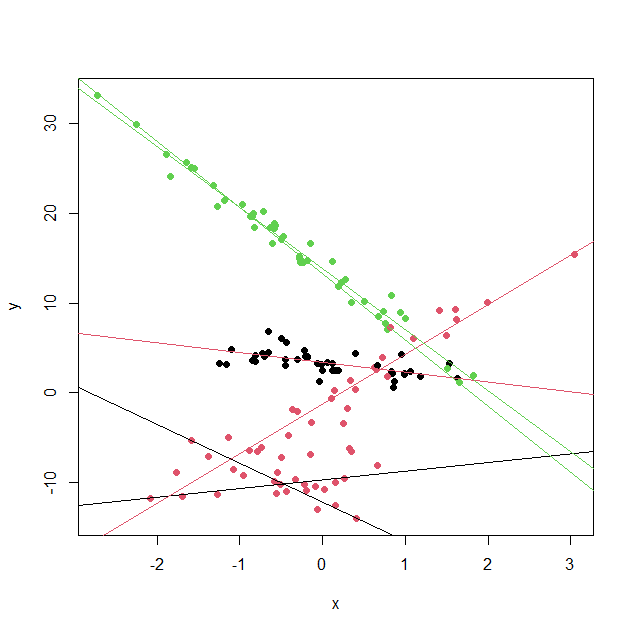}}
\caption{The regime Markov switching example}\label{regress}
\end{figure}

To fit the regime-switching additive regression model to the \code{train} data, we make an
initial clustering of the data, using the \code{initial\_cluster} function, by letting \code{nstate = 3}, \code{nmix = NULL} and \code{regress = TRUE}. Using the argument \code{nmix = NULL} is essential in this case, since the parameters of the regime switching additive regression model does not involve mixture components.
\begin{lstlisting}
|\bf \color{lightgray} R$>$| clus = initial_cluster(train = train, nstate = 3, nmix = NULL,
|\bf \color{lightgray} +|     verbose = FALSE, regress = TRUE)
\end{lstlisting}
Now, we initialize the model using the function \code{initialize\_model}, with arguments \code{mstep = additive\_reg\_mstep} and \code{dens.emission = dnorm\_additive\_reg}. Note that here, we only consider a full-Markovian model and thus we let \code{semi = rep(FALSE, 3)} and \code{sojourn = NULL}, while one can also consider HSMM or HHSMM models by considering different \code{semi} and \code{sojourn} arguments. 
\begin{lstlisting}
|\bf \color{lightgray} R$>$| initmodel = initialize_model(clus = clus ,mstep = additive_reg_mstep,
|\bf \color{lightgray} +|     dens.emission = dnorm_additive_reg, sojourn = NULL, semi = rep(FALSE, 3),
|\bf \color{lightgray} +|     M = max(train$N), verbose = FALSE)
\end{lstlisting}
Next, we fit the model by calling the \code{hhsmmfit} function.
\begin{lstlisting}
|\bf \color{lightgray} R$>$| fit1 = hhsmmfit(x = train, model = initmodel, mstep = additive_reg_mstep,
|\bf \color{lightgray} +|     M = max(train$N), par = list(verbose = FALSE))
\end{lstlisting}
Again, we plot the data to add the fitted lines. The colors of the points show the true states, while the characters present the predicted states.
\begin{lstlisting}
|\bf \color{lightgray} R$>$| plot(train$x[, 1] ~ train$x[, 2], col = train$s, pch = fit1$yhat,
|\bf \color{lightgray} +|     xlab = "x", ylab = "y")
|\bf \color{lightgray} R$>$| text(0, 30, "colors are real states",col="red")
|\bf \color{lightgray} R$>$| text(0, 28, "characters are predicted states")
\end{lstlisting}
To obtain the predicted values of the response variable, we use the \code{addreg\_hhsmm\_predict} function as follows.
\begin{lstlisting}
|\bf \color{lightgray} R$>$| pred <- addreg_hhsmm_predict(fit1, train$x[, 2], 5)
|\bf \color{lightgray} R$>$| yhat1 <- pred[[1]]
|\bf \color{lightgray} R$>$| yhat2 <- pred[[2]]
|\bf \color{lightgray} R$>$| yhat3 <- pred[[3]]
\end{lstlisting}
We add the predicted curves to the plot. The resulting plot is shown in Figure \ref{addregress}.
\begin{lstlisting}
|\bf \color{lightgray} R$>$| lines(yhat1[order(train$x[, 2])]~sort(train$x[, 2]),col = 2)
|\bf \color{lightgray} R$>$| lines(yhat2[order(train$x[, 2])]~sort(train$x[, 2]),col = 1)
|\bf \color{lightgray} R$>$| lines(yhat3[order(train$x[, 2])]~sort(train$x[, 2]),col = 3)
\end{lstlisting}
As one can see from Figure \ref{addregress}, the curves have a proper fit to the data points.
\begin{figure}
\centerline{\includegraphics[scale=0.6]{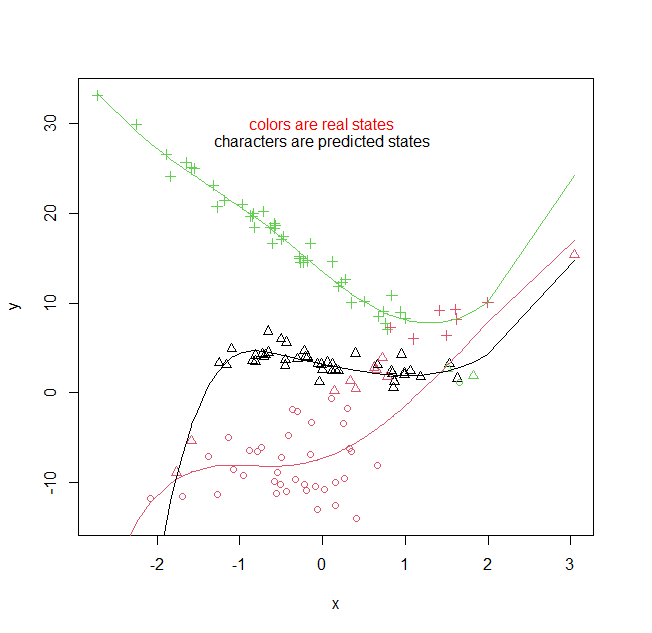}}
\caption{The Markov regime-switching additive regression fit.}\label{addregress}
\end{figure}

\subsection{Auto-regressive HHSMM}

A special case of the regime-switching regression models \eqref{rsrm} and \eqref{rsrm2} is the auto-regressive HHSMM model, in which we take $x_t = (y_{t-1},\ldots,y_{t-\ell})$, for a specified lag $\ell \geq 1$. Here, we present a simulated data example, to illustrate this special case.

The model specification of the auto-regressive HHSMM is similar to that of the regime-switching regression model, noting that the dimension of $x_t$ is always $\ell$ times the dimension of $y_t$. So, we specify the model as follows.

\begin{lstlisting}
|\bf \color{lightgray} R$>$| J <- 2
|\bf \color{lightgray} R$>$| initial <- c(1, 0)
|\bf \color{lightgray} R$>$| semi <- rep(FALSE, 2)
|\bf \color{lightgray} R$>$| P <- matrix(c(0.2, 0.8, 0.1, 0.9),
|\bf \color{lightgray} +|     nrow = J, byrow = TRUE)
|\bf \color{lightgray} R$>$| par <- list(intercept = list(0.5, -0.8),
|\bf \color{lightgray} +|     coefficient = list(-0.8, 0.7),
|\bf \color{lightgray} +|     csigma = list(0.5, 0.2), mix.p = list(1, 1))
|\bf \color{lightgray} R$>$| model <- hhsmmspec(init = initial, transition = P,
|\bf \color{lightgray} +|     parms.emis = par, dens.emis = dmixlm, semi = semi)
\end{lstlisting}

To simulate the data using the \code{simulate.hhsmm} function, we have to use the argument \code{emission.control = list(autoregress = TRUE)}. We, then, plot the simulated data by using \code{plot.hhsmmdata} as follows. The resulting plot is shown in Figure \ref{areg}.

\begin{lstlisting}
|\bf \color{lightgray} R$>$| train <- simulate(model, nsim = c(50, 60, 84, 100),
|\bf \color{lightgray} +|     seed = 1234, emission.control = list(autoregress = TRUE))
|\bf \color{lightgray} R$>$| plot(train)
\end{lstlisting}

\begin{figure}
\centerline{\includegraphics[scale=0.5]{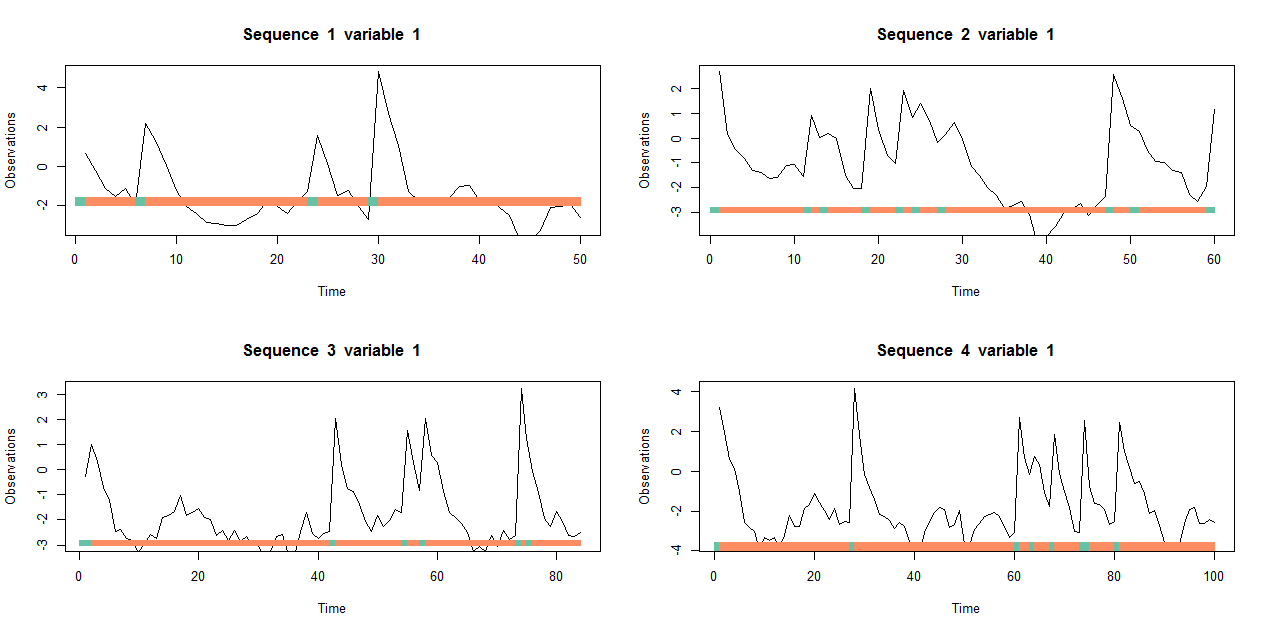}}
\caption{The ARHMM example simulated train data set }\label{areg}
\end{figure}

To prepare the data for fitting the regime-switching regression model, we should first construct the lagged data matrix by using the function \code{lagdata} as follows. The default of the parameter \code{lags} of this function is equal to 1, which is the number of lags to be calculated.

\begin{lstlisting}
|\bf \color{lightgray} R$>$| train2 = lagdata(train)
\end{lstlisting}

The resulting lagged data set is then used for obtaining the initial clustering, using the argument \code{regress = TRUE} and \code{resp.ind = 2} in the \code{initial\_cluster} function as follows.

\begin{lstlisting}
|\bf \color{lightgray} R$>$| clus = initial_cluster(train = train2, nstate = 2, nmix = 1,
|\bf \color{lightgray} +|     ltr = FALSE, final.absorb = FALSE, verbose = FALSE,
|\bf \color{lightgray} +|     regress = TRUE, resp.ind = 2)
\end{lstlisting}

Now, we initialize and fit the model as before. Note that we should use the argument  \code{resp.ind = 2} in place of \code{...} in both functions (see the manual of the \pkg{hhsmm} package \url{https://cran.r-project.org/web/packages/hhsmm/hhsmm.pdf})

\begin{lstlisting}
|\bf \color{lightgray} R$>$| initmodel = initialize_model(clus = clus, mstep = mixlm_mstep,
|\bf \color{lightgray} +|     dens.emission = dmixlm, sojourn = NULL, semi=rep(FALSE, 2),
|\bf \color{lightgray} +|     M = max(train$N), verbose = FALSE, resp.ind = 2)
|\bf \color{lightgray} R$>$| fit1 = hhsmmfit(x = train2, model = initmodel, mstep = mixlm_mstep,
|\bf \color{lightgray} +|     resp.ind = 2, M = max(train$N), par = list(verbose = FALSE))
\end{lstlisting}

To test the performance of the fitted model for prediction of the future time series, we need to simulate a test data set and then right-trim the test data set, using the \code{train\_test\_split}, and by setting \code{train.ratio = 1}, \code{trim = TRUE} and \code{trim.ratio = 0.9}, as follows. As one can see, the length of the sequences in \code{trimmed\_test\$trimmed} is $90\%$ of the associated lengths in \code{test} data set.

\begin{lstlisting}
|\bf \color{lightgray} R$>$| test <- simulate(model, nsim = c(100, 95), seed = 1234,
|\bf \color{lightgray} +|     emission.control = list(autoregress = TRUE))
|\bf \color{lightgray} R$>$| trimmed_test = train_test_split(test, train.ratio = 1,
|\bf \color{lightgray} +|     trim = TRUE, trim.ratio = 0.9)
|\bf \color{lightgray} R$>$| test$N
[1]  100  95
|\bf \color{lightgray} R$>$| trimmed_test$trimmed$N
[1] 90 85
|\bf \color{lightgray} R$>$| trimmed = trimmed_test$trimmed
|\bf \color{lightgray} R$>$| tc = trimmed_test$trimmed.count
\end{lstlisting}
The option \code{train.ratio = 1} means that we do not wish to split the test samples into new train and test subsets and we only need to right trim the sequences.
Now, we have both trimmed sequences in \code{trimmed} object and the complete test samples in \code{test} data set, so that we can compare the true and predicted states. The object \code{tc} contains the number of trimmed items in each sequence, which has to be predicted.

Now, we use the estimated parameters of the ARHMM to predict the future values of the sequence. To do this, we predict the state sequence of the lagged trimmed test data set using the \code{predict.hhsmm} function and then we obtain the linear predictors for the future values as follows.

\begin{lstlisting}
|\bf \color{lightgray} R$>$| lag_trimmed = lagdata(trimmed)
|\bf \color{lightgray} R$>$| n <- length(tc)
|\bf \color{lightgray} R$>$| Ncl <- cumsum(c(0, lag_trimmed$N))
|\bf \color{lightgray} R$>$| Nc2 <- cumsum(c(0, test$N))
|\bf \color{lightgray} R$>$| Nc <- cumsum(c(0, trimmed$N))
|\bf \color{lightgray} R$>$| par(mfrow = c(2,1))
|\bf \color{lightgray} R$>$| for (i in 1:n) {
|\bf \color{lightgray} +|     new_lag_data = list(x = lag_trimmed$x[(Ncl[i] + 1):Ncl[i + 1], ],
|\bf \color{lightgray} +|         N = lag_trimmed$N[i])
|\bf \color{lightgray} +|     new_data = list(x = as.matrix(trimmed$x[(Nc[i] + 1):Nc[i + 1], ]),
|\bf \color{lightgray} +|         N = trimmed$N[i])
|\bf \color{lightgray} +|     yhat1 <- predict(fit1, new_lag_data, future = tc[i])
|\bf \color{lightgray} +|     fstates <- yhat1$s[((test$N[i] - tc[i])):(test$N[i] - 1)]
|\bf \color{lightgray} +|     intercept = coefficients = csigma = c()
|\bf \color{lightgray} +|     xcurrent = as.vector(new_data$x[new_data$N, ])
|\bf \color{lightgray} +|     pred <- xcurrent
|\bf \color{lightgray} +|     for(j in 1:tc[i]){
|\bf \color{lightgray} +|         intercept[j] <-
|\bf \color{lightgray} +|             as.vector(fit1$model$parms.emission$intercept[[fstates[j]]])
|\bf \color{lightgray} +|         coefficients[j] <-
|\bf \color{lightgray} +|             as.vector(fit1$model$parms.emission$coefficients[[fstates[j]]])
|\bf \color{lightgray} +|         predicted <- intercept[j] + xcurrent * coefficients[j]
|\bf \color{lightgray} +|         xcurrent <- predicted
|\bf \color{lightgray} +|         pred <- c(pred, predicted)
|\bf \color{lightgray} +|     }
|\bf \color{lightgray} +|     tr_time = ((test$N[i] - tc[i] - 1)):(test$N[i] - 1) + 1
|\bf \color{lightgray} +|     plot(test$x[(Nc2[i] + 1):Nc2[i + 1], ], type = "l", xlab = "time",
|\bf \color{lightgray} +|         ylab = "x", main = paste("sequence", i))
|\bf \color{lightgray} +|     lines(pred ~ tr_time, lwd = 3, col = 2)
|\bf \color{lightgray} +| }
\end{lstlisting}

The resulting plot is presented in Figure \ref{pred}. The colored lines are the predicted values.

\begin{figure}
\centerline{\includegraphics[scale=0.5]{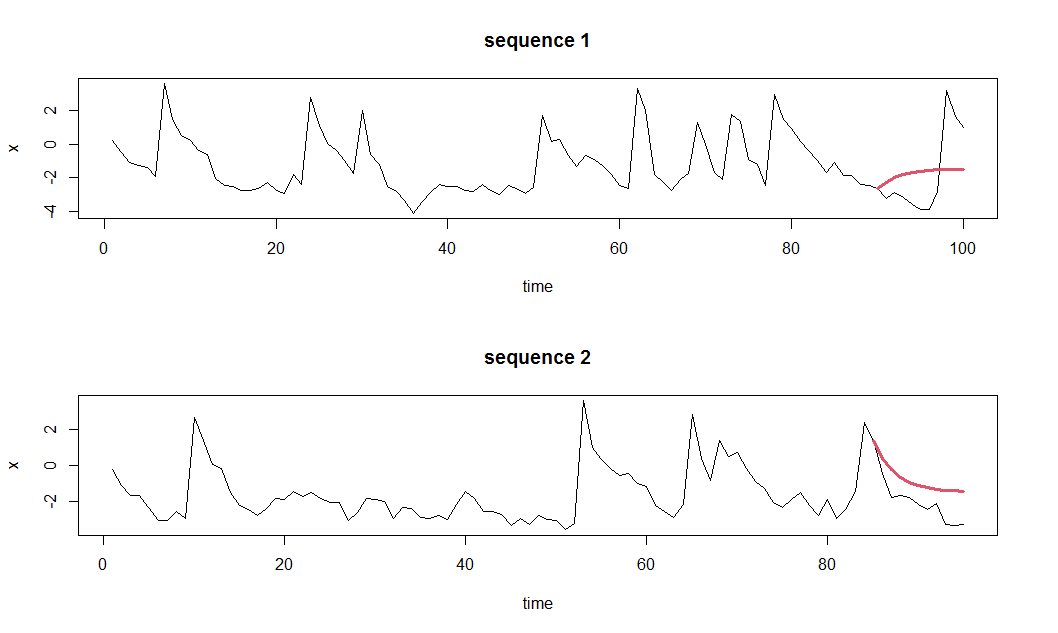}}
\caption{Trimmed test data set and the predicted values}\label{pred}
\end{figure}

We try to apply the regime-switching additive regression model to fit the AR-HMM model. Again, the only
difference of the initial clustering using the \code{initial\_cluster} function is to set \code{nmix = NULL}.

\begin{lstlisting}
|\bf \color{lightgray} R$>$| clus = initial_cluster(train = train2, nstate = 2, nmix = NULL,
|\bf \color{lightgray} +|     verbose = FALSE, regress = TRUE, resp.ind = 2)
\end{lstlisting}
We initialize the model using the function \code{initialize\_model} by setting \code{mstep = additive\_reg\_mstep} and  \code{dens.emission = dnorm\_additive\_reg}. The difference here is that we pass the parameters of these function
to the \code{initialize\_model} function through the argument \code{control}. In the following, we pass the response indicator and the degree of the B-splines by setting the argument \code{control = list(resp.ind = 2, K = 7)}.
\begin{lstlisting}
|\bf \color{lightgray} R$>$| initmodel = initialize_model(clus = clus, mstep =
|\bf \color{lightgray} +|     additive_reg_mstep, dens.emission = dnorm_additive_reg,
|\bf \color{lightgray} +|     sojourn = NULL, semi = rep(FALSE, 2), M = max(train$N),
|\bf \color{lightgray} +|     verbose = FALSE, control = list(resp.ind = 2, K = 7))
\end{lstlisting}
Now, we fit the model by calling the \code{hhsmmfit} function, as follows.
\begin{lstlisting}
|\bf \color{lightgray} R$>$| fit1 = hhsmmfit(x = train2, model = initmodel,
|\bf \color{lightgray} +|     mstep = additive_reg_mstep, M = max(train$N),
|\bf \color{lightgray} +|     control = list(resp.ind = 2, K = 7),
|\bf \color{lightgray} +|     par = list(verbose = FALSE))
\end{lstlisting}
Finally, we provide the prediction plots, through the following codes.
\begin{lstlisting}
|\bf \color{lightgray} R$>$| par(mfrow = c(2,1))
|\bf \color{lightgray} R$>$| for (i in 1:n) {
|\bf \color{lightgray} +|      new_lag_data = list(x = lag_trimmed$x[(Ncl[i] + 1):Ncl[i + 1], ],
|\bf \color{lightgray} +|          N = lag_trimmed$N[i])
|\bf \color{lightgray} +|      new_data = list(x = as.matrix(trimmed$x[(Nc[i] + 1):Nc[i + 1], ]),
|\bf \color{lightgray} +|          N = trimmed$N[i])
|\bf \color{lightgray} +|      yhat1 <- predict(fit1, new_lag_data, future = tc[i])
|\bf \color{lightgray} +|      fstates <- yhat1$s[((test$N[i] - tc[i])):(test$N[i] - 1)]
|\bf \color{lightgray} +|      tr_time = ((test$N[i] - tc[i] - 1)):(test$N[i] - 1) + 1
|\bf \color{lightgray} +|      intercept =  c()
|\bf \color{lightgray} +|      coefficients = list()
|\bf \color{lightgray} +|      xcurrent = as.vector(new_data$x[new_data$N, ])
|\bf \color{lightgray} +|      pred <- xcurrent
|\bf \color{lightgray} +|      for(j in 1:tc[i]){
|\bf \color{lightgray} +|         preds <- addreg_hhsmm_predict(fit1, c(xcurrent,train$x), 7)
|\bf \color{lightgray} +|         predicted <- preds[[fstates[j]]][1]
|\bf \color{lightgray} +|         xcurrent <- predicted
|\bf \color{lightgray} +|         pred <- c(pred, predicted)
|\bf \color{lightgray} +|     }
|\bf \color{lightgray} +|     plot(test$x[(Nc2[i] + 1):Nc2[i + 1], ], type = "l", xlab = "time",
|\bf \color{lightgray} +|     ylab = "x", main = paste("sequence", i))
|\bf \color{lightgray} +|     lines(pred ~ tr_time, lwd = 3, col = 2)
|\bf \color{lightgray} +| }
\end{lstlisting}
The resulting plots are presented in Figure \ref{pred2}. The colored lines are the predicted values. By comparing the Figures \ref{pred} and \ref{pred2} one can see that the regime-switching additive regression model results in more accurate prediction especially for the second sequence of the \code{test} data set.

\begin{figure}
\centerline{\includegraphics[scale=0.5]{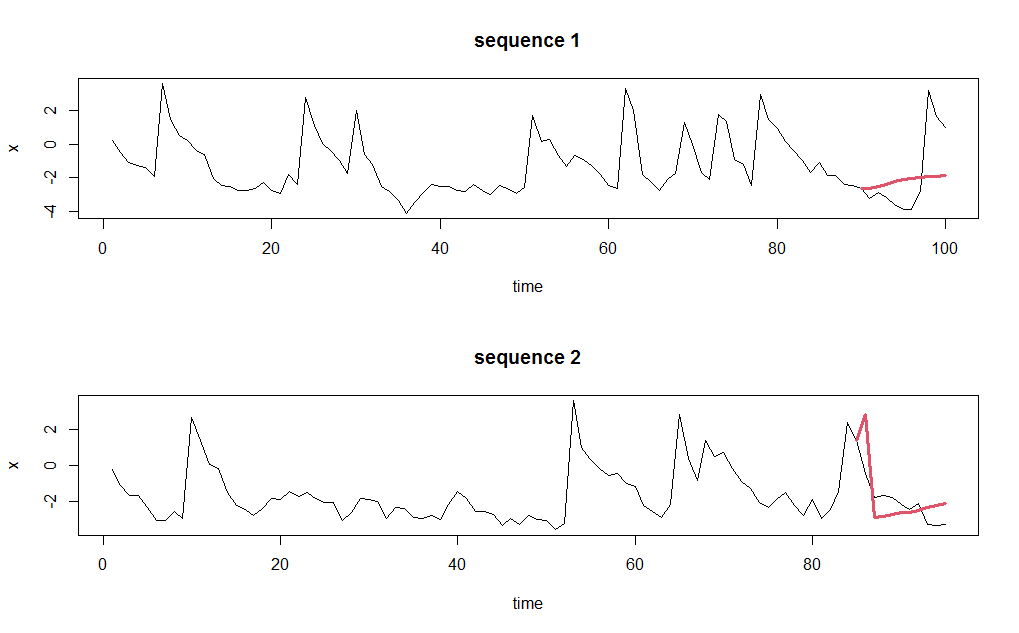}}
\caption{The predicted values using AR-HMM model, using the regime switching additive regression}\label{pred2}
\end{figure}

}}

{\subsection{Prediction of the future state sequence}

To predict the future state sequence at times $T+1,\ldots,T+h$, first, we use viterbi (smoothing) algorithm (see the Appendix) to estimate the probabilities of the most likely path $\alpha_j(t)$ ($L_j(t)$) for $j=1,\ldots,J$ and $t=0,\ldots,\tau-1$, as well as the current most likely state $\hat{s}_t^* =\arg\max_{1\leq j\leq J} \alpha_j(t)$ ($\hat{s}_t^* =\arg\max_{1\leq j\leq J} L_j(t)$). Also, we might compute the probabilities
\begin{equation}\label{deltabar}
\bar{\delta}_t(j)=\frac{\alpha_{j}(t)}{\sum_{k=1}^J\alpha_{k}(t)}\quad  (\bar{\delta}_t(j)=\frac{L_{j}(t)}{\sum_{k=1}^JL_{k}(t)}).
\end{equation}
Next, for $j=1,\ldots,h$, we compute the probability of the next state, by multiplying the transition matrix by the current state probability as follows
\begin{equation}\label{52__eq2}
\bar{\delta}_{next} = \Big(P\Big)^T\bar{\delta}_{current}
\end{equation}
Then, the $j$th future state  is predicted as
\begin{equation}\label{53__eq}
\hat{s}_{next}^*=\arg\max_{1\leq j \leq J}\bar{\delta}_{next}(j)
\end{equation}
This process continues until the required time $T+h$. The prediction of the future state sequence is done using the function \code{predict.hhsmm} function in the \pkg{hhsmm} package, by determining the argument \code{future}, which is equal to zero by default. To examine this ability, we study a simple example as follows. First, we define a simple model, just like the model in section 3, and simulate train and test samples from this model, as follows.
\begin{lstlisting}
|\bf \color{lightgray} R$>$| J <- 3
|\bf \color{lightgray} R$>$| initial <- c(1, 0, 0)
|\bf \color{lightgray} R$>$| semi <- c(FALSE, TRUE, FALSE)
|\bf \color{lightgray} R$>$| P <- matrix(c(0.8, 0.1, 0.1, 0.5, 0, 0.5, 0.1, 0.2, 0.7),
|\bf \color{lightgray} +|     nrow = J, byrow = TRUE)
|\bf \color{lightgray} R$>$| par <- list(mu = list(list(7, 8),list(10, 9, 11),list(12, 14)),
|\bf \color{lightgray} +|     sigma = list(list(3.8, 4.9),list(4.3, 4.2, 5.4),list(4.5, 6.1)),
|\bf \color{lightgray} +|     mix.p = list(c(0.3, 0.7),c(0.2, 0.3, 0.5),c(0.5, 0.5)))
|\bf \color{lightgray} R$>$| sojourn <- list(shape = c(0, 3, 0), scale = c(0, 10, 0),
|\bf \color{lightgray} +|     type = "gamma")
|\bf \color{lightgray} R$>$| model <- hhsmmspec(init = initial, transition = P,
|\bf \color{lightgray} +|     parms.emis = par, dens.emis = dmixmvnorm,
|\bf \color{lightgray} +|     sojourn = sojourn, semi = semi)
|\bf \color{lightgray} R$>$| train <- simulate(model, nsim = c(50, 40, 30, 70), seed = 1234,
|\bf \color{lightgray} +|     remission = rmixmvnorm)
|\bf \color{lightgray} R$>$| test <- simulate(model, nsim = c(80, 45, 20, 35), seed = 1234,
|\bf \color{lightgray} +|     remission = rmixmvnorm)
\end{lstlisting}
To examine the prediction performance of the model, we split the test sample from the right, using \code{train\_test\_split} function and a trim ratio equal to 0.9, as follows. 
\begin{lstlisting}
|\bf \color{lightgray} R$>$| tt = train_test_split(test, train.ratio = 1, trim = TRUE,
|\bf \color{lightgray} +|     trim.ratio = 0.9)
|\bf \color{lightgray} R$>$| trimmed = tt$trimmed
|\bf \color{lightgray} R$>$| tc = tt$trimmed.count
\end{lstlisting}
As in Section \ref{s3}, we initialize and fit an HHSMM model to the \code{train} data set, as follows.
\begin{lstlisting}
|\bf \color{lightgray} R$>$| clus = initial_cluster(train, nstate=3, nmix=c(2, 2, 2),
|\bf \color{lightgray} +|     ltr = FALSE, final.absorb = FALSE, verbose = FALSE)
|\bf \color{lightgray} R$>$| semi <- c(FALSE, TRUE, FALSE)
|\bf \color{lightgray} R$>$| initmodel1 = initialize_model(clus = clus, sojourn = "gamma",
|\bf \color{lightgray} +|     M = max(train$N), semi = semi, verbose = FALSE)
|\bf \color{lightgray} R$>$| fit1 = hhsmmfit(x = train, model = initmodel1, M = max(train$N),
|\bf \color{lightgray} +|     par = list(verbose = FALSE))
\end{lstlisting}
Now, we predict the future states of each sequence of the test data set, separately, using the option \code{future = tc[i]}. Then, we print the homogeneity of real and predicted state sequences, by using the \code{homogeneity} function, as follows.
\begin{lstlisting}
|\bf \color{lightgray} R$>$| n <- length(tc)
|\bf \color{lightgray} R$>$| Nc <- cumsum(c(0, trimmed$N))
|\bf \color{lightgray} R$>$| Nc2 <- cumsum(c(0, test$N))
|\bf \color{lightgray} R$>$| for(i in 1:n){
|\bf \color{lightgray} +|     newdata = list(x = trimmed$x[(Nc[i] + 1):Nc[i + 1], ],
|\bf \color{lightgray} +|         N = trimmed$N[i])
|\bf \color{lightgray} +|     yhat1 <- predict(fit1, newdata, future = tc[i])
|\bf \color{lightgray} +|     yhat2 <- predict(fit1, newdata, future = 0)
|\bf \color{lightgray} +|     cat("homogeneity with future sequence")
|\bf \color{lightgray} +|     print(homogeneity(yhat1$s , test$s[(Nc2[i] + 1):Nc2[i + 1]]))
|\bf \color{lightgray} +|     cat("homogeneity without future sequence")
|\bf \color{lightgray} +|     print(homogeneity(yhat2$s , trimmed$s[(Nc[i] + 1):Nc[i + 1]]))
|\bf \color{lightgray} +| }

homogeneity with future sequence[1] 0.8965517 0.8066667 0.6097561
homogeneity without future sequence[1] 0.9079755 0.8066667 0.6097561
homogeneity with future sequence[1] 0.8205128 0.9000000 0.0000000
homogeneity without future sequence[1] 0.9140625 0.9000000 0.0000000
homogeneity with future sequence[1] 1 1 1
homogeneity without future sequence[1] 1 1 1
homogeneity with future sequence[1] 0.9333333 0.8306452 0.6578947
homogeneity without future sequence[1] 0.8750000 0.8306452 0.6578947
\end{lstlisting}
As, on can see from the above homogeneities, the predictions are quite good.
}

\subsection{Residual useful lifetime (RUL) estimation, for reliability applications}\label{s4}

The residual useful lifetime (RUL) is defined as the remaining lifetime of a system at a specified time point. If we analyse a reliable system with a hidden Markov or semi-Markov model, a suitable choice would be a left to right model, with the final state as the failure state. The RUL of such model is defined
at time $t$ as
\begin{eqnarray}\label{47__eq}
\mbox{RUL}_t = \tilde{D} :S_{t+\tilde{D}} = J, S_{t+\tilde{D}-1} = i;\hspace{1cm}1\leq i < k\leq J.
\end{eqnarray}
We describe a method of RUL estimation \citep[see][]{cea15}, which is used in \pkg{hhsmm} package. 
First, we should compute the probabilities in \eqref{deltabar}, using the Viterbi or smoothing algorithm. 

Two different methods are used to obtain point and interval estimates of the RUL in the \pkg{hhsmm} package. The first method (the option \code{confidence = "mean"} in the \code{predict} function of the \pkg{hhsmm} package) is based on the method described in \cite{cea15}. This method computes the average time in the current state as follows
\begin{eqnarray}\label{49__eq}
\tilde{d}_{avg}(\hat{s}_t^*) = \sum_{j=1}^{J}\Big(\mu_{d_j}-\hat{d}_t(j)\Big)\bar{\delta}_t(j),
\end{eqnarray}
where  $\mu_{d_i}=\sum_{u=1}^{M_j} u d_i(u)$ is the expected value of the duration variable in state $j$, and  $\hat{d}_t(j)$
is the estimated states duration, computed as follows \citep{a04}
$$\hat{d}_t(j) = \hat{d}_{t-1}(j) \bar{\delta}_t(j), \quad t = 2, \ldots , M_j, \quad  \hat{d}_{1}(j)=1, \; j = 1,\ldots, J. $$
In order to obtain a confidence interval for the RUL, \cite{cea15} also computed the standard deviation of the duration variable in state $j$, $\sigma_{d_j}$, and
\begin{equation}\label{50__eq}
\tilde{d}_{low}(\hat{s}_t^*) = \sum_{j=1}^{J}\Big(\mu_{d_j}-\hat{d}_t(j)-\sigma_{d_j}\Big)\bar{\delta}_t(j),
\end{equation}
\begin{equation}\label{51__eq}
\tilde{d}_{up}(\hat{s}_t^*) =\sum_{j=1}^{J}\Big(\mu_{d_j}-\hat{d}_t(j)+\sigma_{d_j}\Big)\bar{\delta}_t(j)
\end{equation}
However, to obtain a confidence interval of the specified level $1-\gamma \in (0,1)$, we have corrected equations \eqref{50__eq} and \eqref{51__eq} in the \pkg{hhsmm} package as follows
\begin{equation}\label{502__eq}
\tilde{d}_{low}(\hat{s}_t^*) = \sum_{j=1}^{J}\Big(\mu_{d_j}-\hat{d}_t(j)-z_{1-\gamma/2}\sigma_{d_j}\Big)\bar{\delta}_t(j),
\end{equation}
\begin{equation}\label{512__eq}
\tilde{d}_{up}(\hat{s}_t^*) =\sum_{j=1}^{J}\Big(\mu_{d_j}-\hat{d}_t(j)+z_{1-\gamma/2}\sigma_{d_j}\Big)\bar{\delta}_t(j),
\end{equation}
where $z_{1-\gamma/2}$ is the ${1-\gamma/2}$ quantile of the standard normal distribution.

The probability of the next state is obtained by multiplying
the transition matrix by the current state probability as follows
\begin{equation}\label{52__eq}
\bar{\delta}_{next} = \Big[\bar{\delta}_{t+\tilde{d}}(j)\Big]_{1\leq j\leq J}=\Big(P\Big)^T\bar{\delta}_t
\end{equation}
while the maximum a posteriori estimate of the next state is calculated as
\begin{equation}\label{53__eq}
\hat{s}_{next}^*=\hat{s}_{t+\tilde{d}}^*=\arg\max_{1\leq j \leq J}\bar{\delta}_{t+\tilde{d}}(j)
\end{equation}
If
$\hat{S}_{t+\tilde{d}}(j)$ coincides with the failure state $J$, the failure
will happen after the remaining time at the current state is
over and the average estimation of the failure time is
 $\tilde{D}_{avg}=\tilde{d}_{avg}(\hat{s_t}^*)$, with the lower and upper bounds $\tilde{D}_{low}=\tilde{d}_{low}(\hat{s_t}^*)$ and  $\tilde{D}_{up}=\tilde{d}_{up}(\hat{s_t}^*)$, respectively, otherwise, the sojourn time of the next state is calculated as
\begin{equation}\label{54__eq}
\tilde{d}_{avg}\Big(\hat{S}_{t+\tilde{d}}^*\Big)=\sum_{j=1}^{J}\mu_{d_j}\bar{\delta}_{t+\tilde{d}}(j)
\end{equation}
\begin{equation}\label{55__eq}
\tilde{d}_{low}\Big(\hat{S}_{t+\tilde{d}}^*\Big)=\sum_{j=1}^{J}\Big(\mu_{d_j}-z_{1-\gamma/2}\sigma_{d_j}\Big)\bar{\delta}_{t+\tilde{d}}(j)
\end{equation}
\begin{equation}\label{56__eq}
\tilde{d}_{up}\Big(\hat{S}_{t+\tilde{d}}^*\Big)=\sum_{j=1}^{J}\Big(\mu_{d_j}+z_{1-\gamma/2}\sigma_{d_j}\Big)\bar{\delta}_{t+\tilde{d}}(j)
\end{equation}
This procedure is iterated until the failure state is
encountered in the prediction of the next state. The estimate
of the RUL is then calculated by summing all the aforementioned
estimated remaining times, as follows
\begin{eqnarray}\label{57__eq}
\tilde{D}_{avg} = \sum\tilde{d}_{avg}, \quad
\tilde{D}_{low} = \sum\tilde{d}_{low}, \quad
\tilde{D}_{up} = \sum\tilde{d}_{up}
\end{eqnarray}
In the second method (the option \code{confidence = "max"} in the \code{predict} function of the \pkg{hhsmm} package), we relax the normal assumption and use the mode and quantiles of the sojourn time distribution, by replacing the mean $\mu_{d_j}$ with the mode $m_{d_j} = \arg\max_{1\leq u \leq M_j} d_j(u)$ and replacing
$-z_{1-\gamma/2}\sigma_{d_j}$ with $\min\{\nu;\; \sum_{u=1}^\nu d_j(u) \leq \gamma/2\}$ and $+z_{1-\gamma/2}\sigma_{d_j}$ with $M_j - \min\{\nu;\; \sum_{u=\nu}^{M_j} d_j(u) \leq \gamma/2\}$  in equations \eqref{50__eq}, \eqref{502__eq}, \eqref{512__eq}, \eqref{54__eq}, \eqref{55__eq} and \eqref{56__eq}.

\subsection{Continuous time sojourn distributions}\label{cts}
Since the measurements of the observations are always preformed on discrete time units (assumed to be positive integers), the sojourn time probabilities of the sojourn time distribution with probability density function $g_j$, in state $j$, is obtained as follows
\begin{eqnarray}\label{cont}
d_j(u) &=& P(S_{t+u+1} \neq j,\;S_{t+u-\nu} = j, \;\nu = 0, \ldots,u-2|S_{t+1} = j ,\; S_t \neq j)\nonumber\\
&=&  \int_{u-1}^{u} g_j(y) \; dy \big/ \int_{0}^{M_j} g_j(y) \; dy, \quad j=1,\ldots,J,\quad u=1,\ldots,M_j.
\end{eqnarray}

{ Almost all flexible continuous distributions with positive domain, which are used as the lifetime distribution, including gamma, weibull, log-normal,  Birnbaum–Saunders, inverse-gamma, Fr\'{e}chet, Gumbel and many other distributions, might be used as the continuous-time sojourn distribution.}  Some of the continuous sojourn time distributions, included in the \pkg{hhsmm} package, are as follows:

\begin{itemize}

\item \textbf{Gamma sojourn}: The gamma sojourn time density functions are

$$g_j(y) = \frac{y^{\alpha_j-1}e^{-\frac{y}{\beta_j}}}{\beta_j^{\alpha_j}\Gamma(\alpha_j)}, \quad j=1,\ldots,J,$$
which result in
$$d_j(u) =  \int_{u-1}^{u} y^{\alpha_j-1}e^{-\frac{y}{\beta_j}} \; dy
 \big/ \int_{0}^{M_j} y^{\alpha_j-1}e^{-\frac{y}{\beta_j}} \; dy
$$

\item \textbf{Weibull sojourn}: The Weibull sojourn time density functions are

$$g_j(y) = \frac{\alpha_j}{\beta_j} \left(\frac{y}{\beta_j}\right)^{\alpha_j-1} \exp\left\{- \left(\frac{y}{\beta_j}\right)^{\alpha_j}\right\}, \quad j=1,\ldots,J,$$
which result in
$$d_j(u) =  \int_{u-1}^{u} y^{\alpha_j-1} \exp\left\{- \left(\frac{y}{\beta_j}\right)^{\alpha_j}\right\} \; dy
 \left./ \int_{0}^{M_j} y^{\alpha_j-1} \exp\left\{- \left(\frac{y}{\beta_j}\right)^{\alpha_j}\right\} \; dy\right.
$$

\item \textbf{log-normal sojourn}: The log-normal sojourn time density functions are

$$g_j(y) = \frac{1}{\sqrt{2\pi}\sigma_j} \exp\left\{\frac{-1}{2\sigma_j^2}(\log y - \mu_j)^2\right\}, \quad j=1,\ldots,J,$$
which result in
$$d_j(u) =  \int_{u-1}^{u} \exp\left\{\frac{-1}{2\sigma_j^2}(\log y - \mu_j)^2\right\} \; dy
 \left./ \int_{0}^{M_j}   \exp\left\{\frac{-1}{2\sigma_j^2}(\log y - \mu_j)^2\right\}  \; dy\right.
$$

\end{itemize}

\subsection{Other features of the package}

There are some other features included in the \pkg{hhsmm} package, which are listed below:

\begin{itemize}
\item \code{dmixmvnorm}: Computes the probability density function of a mixture of multivariate normals
for a specified observation vector, a specified state, and a specified model's parameters
\item \code{mixmvnorm\_mstep}: The M step function of the EM algorithm for the mixture
of multivariate normals as the emission distribution using the
observation matrix and the estimated weight vectors
\item \code{rmixmvnorm}: Generates a vector of observations from mixture multivariate
 normal distribution in a specified state and using the  parameters of a specified model
\item \code{train\_test\_split}:  Splits the data sets to train and test
subsets with an option to right trim the sequences
\item \code{lagdata}: Creates lagged time series of a data
\item \code{score}: Computes the score (log-likelihood) of new observations using a trained model
\item \code{homogeneity}:  Computes maximum homogeneity of two state sequences
\item \code{hhsmmdata}: Converts a matrix of data and its associated vector  of sequence lengths to a data list of class \code{"hhsmmdata"}
\end{itemize}

{
\section{Real data Analysis}\label{rdas}

To examine the performance of the \pkg{hhsmm} package, we consider the analysis of two real data sets. The first data set is the Spain energy market data set from \pkg{MSwM} package and the second one is the  Commercial Modular Aero-Propulsion System Simulation (CMAPSS) data set from the \pkg{CMAPSS} data package.

\subsection{Spain energy market data set}

The Spain energy market data set \citep{fea09} contains the price of the energy in Spain with other economic data. The daily data is from January 1, 2002 to October 31, 2008, during working days (Monday to Friday). 
This data set is available in \pkg{MSwM} package (\url{https://cran.r-project.org/package=MSwM}), in a data-frame named \code{energy} and contains 1785 observations on 7 variables: \code{Price} (Average price of energy in Cent/kwh), 
\code{Oil} (Oil price in Euro/barril), \code{Gas} (Gas price in Euro/MWh), \code{Coal} (Coal price in Euro/T), \code{EurDol} (Exchange rate between Dolar-Euro in USD-Euro), \code{Ibex35} (Ibex 35 index divided by one thousand) and \code{Demand} (Daily demand of energy in GWh). This data-set is also analysed in \cite{fea09}, using Markov switching regression model. The objective of the analysis is to predict the response variable \code{Price}, based on the information in other variables (covariates). 

In order to analyze the \code{energy} data set, we load it from \pkg{MSwM} package, and transform it into a \code{"hhsmmdata"} using \code{hhsmmdata} function as follows.

\begin{lstlisting}
|\bf \color{lightgray} R$>$| library(MSwM)
|\bf \color{lightgray} R$>$| data(energy)
|\bf \color{lightgray} R$>$| energy.hmm = hhsmmdata(energy)
|\bf \color{lightgray} R$>$| p = ncol(energy.hmm$x) - 1
\end{lstlisting}

We consider a two-state model. Here, we consider a fully Markovian model and thus, we let \code{semi <- rep(FALSE, J)}. 
Although an optimal value of $K$ might be obtained by minimizing the AIC, BIC or even a cross-validation error, we set the degree of the B-splines to $K=20$ for this analysis, 
for the sake of briefness. 

\begin{lstlisting}
|\bf \color{lightgray} R$>$| K <- 20
|\bf \color{lightgray} R$>$| J <- 2
|\bf \color{lightgray} R$>$| initial <- rep(1/J,  J)
|\bf \color{lightgray} R$>$| semi <- rep(FALSE, J)
\end{lstlisting}

First, we make an initial clustering for the data set. Again, we point out that we should consider \code{nmix = NULL} and \code{regress = TRUE} in the \code{initial\_cluster} function.

\begin{lstlisting}
|\bf \color{lightgray} R$>$| clus = initial_cluster(train = energy.hmm, nstate = J,
|\bf \color{lightgray} +|     nmix = NULL, ltr = FALSE, final.absorb = FALSE,
|\bf \color{lightgray} +|     verbose = TRUE, regress = TRUE)
\end{lstlisting}

To initialize the model, we use the \code{initialize\_model} function, with arguments \code{mstep = additive\_reg\_mstep}, \code{dens.emission = dnorm\_additive\_reg} and \code{control = list(K = K)}.

\begin{lstlisting}
|\bf \color{lightgray} R$>$| initmodel = initialize_model(clus = clus, mstep =
|\bf \color{lightgray} +|     additive_reg_mstep, dens.emission = dnorm_additive_reg,
|\bf \color{lightgray} +|     sojourn = NULL, semi = semi, M = max(energy.hmm$N),
|\bf \color{lightgray} +|     verbose = FALSE, control = list(K = K))
\end{lstlisting}

Next, we fit the model by calling the \code{hhsmmfit} function as follows.

\begin{lstlisting}
|\bf \color{lightgray} R$>$| fit1 = hhsmmfit(x = energy.hmm, model = initmodel,
|\bf \color{lightgray} +|     mstep = additive_reg_mstep, M = max(energy.hmm$N),
|\bf \color{lightgray} +|     control = list(K = K))
\end{lstlisting}

Now, we can obtain the response predictions using the \code{addreg\_hhsmm\_predict} function as follows.

\begin{lstlisting}
|\bf \color{lightgray} R$>$| pred2 <- addreg_hhsmm_predict(fit1, energy.hmm$x[, 2:(p+1)], K)
\end{lstlisting}

To visualize the results, first, we add the predicted states to a \code{hhsmmdata} set made by the response variable. Then, we plot it using the \code{plot.hhsmmspec} function, as follows.

\begin{lstlisting}
|\bf \color{lightgray} R$>$| s = fit1$yhat
|\bf \color{lightgray} R$>$| newdata = hhsmmdata(x = as.matrix(energy.hmm$x[, 1]),
|\bf \color{lightgray} +|     N = energy.hmm$N)
|\bf \color{lightgray} R$>$| newdata$s = s
|\bf \color{lightgray} R$>$| plot(newdata)
\end{lstlisting}

The resulting plot is presented in Figure \ref{spain1}. The predicted states are shown by two different colors on the horizontal axis. 

\begin{figure}
\centerline{\includegraphics[width = 15cm]{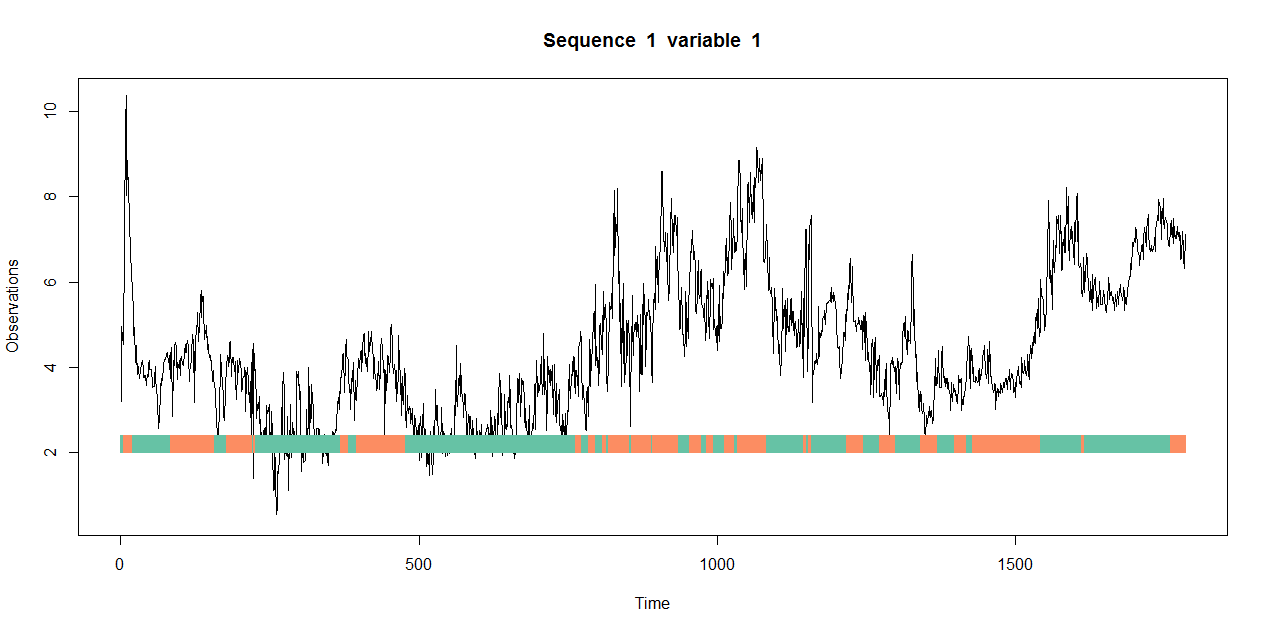}}
\caption{Spain energy data and its estimated states.}\label{spain1}
\end{figure}

In the second plot, we want to show the separate predictions for the two states along with the true response values. To do this, we use the following lines of codes.

\begin{lstlisting}
|\bf \color{lightgray} R$>$| col.states = fit1$yhat
|\bf \color{lightgray} R$>$|  col.states[col.states == 1] = 'goldenrod2'
|\bf \color{lightgray} R$>$|  col.states[col.states == 2] = 'green4'
|\bf \color{lightgray} R$>$| plot(energy.hmm$x[, 1], col = col.states,
|\bf \color{lightgray} +|     pch = 16, xlab = "Time", ylab = "Energy Price")
|\bf \color{lightgray} R$>$| time = 1:length(pred2[[1]])
|\bf \color{lightgray} R$>$| lines(pred2[[1]] ~ time, col = "red", lwd = 0.25)
|\bf \color{lightgray} R$>$| lines(pred2[[2]] ~ time, col = "blue", lwd = 0.25)
\end{lstlisting}

The resulting plot is presented in Figure \ref{spain2}. The predictions associated with the two states' are shown by blue and red lines.

\begin{figure}
\centerline{\includegraphics[width = 15cm]{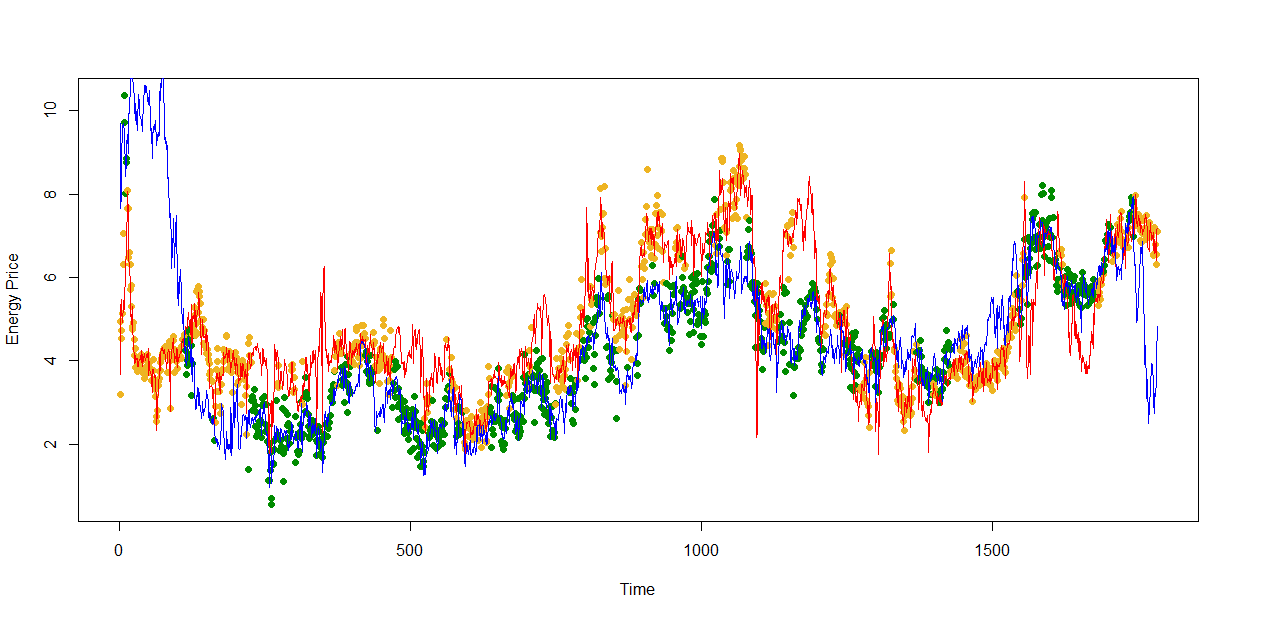}}
\caption{Two-state prediction of the energy price based on the other variables.}\label{spain2}
\end{figure}

To visualize the nonparametric regression curves, we consider only the covariate \code{Oil Price}, which is the second column of the \code{energy} data set. In the following, we initialize and fit the model to the data set containing only the first column as the response and the second column as the covariate.

\begin{lstlisting}
|\bf \color{lightgray} R$>$| energy.hmm2 = hhsmmdata(energy[ , 1:2])
|\bf \color{lightgray} R$>$| clus = initial_cluster(train = energy.hmm2,  nstate = J, nmix = NULL,
|\bf \color{lightgray} +|     regress = TRUE)
|\bf \color{lightgray} R$>$| initmodel = initialize_model(clus = clus, mstep = additive_reg_mstep,
|\bf \color{lightgray} +|     dens.emission = dnorm_additive_reg, sojourn = NULL, semi = semi,
|\bf \color{lightgray} +|     M = max(energy.hmm$N))
|\bf \color{lightgray} R$>$| fit1 = hhsmmfit(x = energy.hmm2, model = initmodel,
|\bf \color{lightgray} +|     mstep = additive_reg_mstep, M = max(energy.hmm2$N))
\end{lstlisting}
Again, we obtain the response predictions as follows.
\begin{lstlisting}
|\bf \color{lightgray} R$>$| pred <- addreg_hhsmm_predict(fit1, energy.hmm2$x[, 2], K)
\end{lstlisting}
Now, we can plot the corresponding graph as follows.
\begin{lstlisting}
|\bf \color{lightgray} R$>$| col.states = fit1$yhat
|\bf \color{lightgray} R$>$| col.states[col.states == 1] = 'goldenrod2'
|\bf \color{lightgray} R$>$| col.states[col.states == 2] = 'green4'
|\bf \color{lightgray} R$>$| plot(energy.hmm$x[, 1] ~ energy.hmm$x[, 2], col = col.states,
|\bf \color{lightgray} +|     pch = 16, xlab = "Oil Price", ylab = "Energy Price", lwd = 2,
|\bf \color{lightgray} +|     main = "Case p = 1")
|\bf \color{lightgray} R$>$| lines(pred[[1]][order(energy.hmm$x[, 2])] ~ sort(energy.hmm$x[, 2]),
|\bf \color{lightgray} +|     col = 1, lwd = 2)
|\bf \color{lightgray} R$>$| lines(pred[[2]][order(energy.hmm$x[, 2])] ~ sort(energy.hmm$x[, 2]),
|\bf \color{lightgray} +|     col = 1, lwd = 2)
|\bf \color{lightgray} R$>$| text(30, 7, "State 1", cex = 1.5)
|\bf \color{lightgray} R$>$| text(60, 3, "State 2", cex = 1.5)
\end{lstlisting}

The resulting plot is presented in Figure \ref{spain3}. As one can see from Figure \ref{spain3}, the two curves are well-fitted to the data points. 

\begin{figure}
\centerline{\includegraphics[width = 15cm]{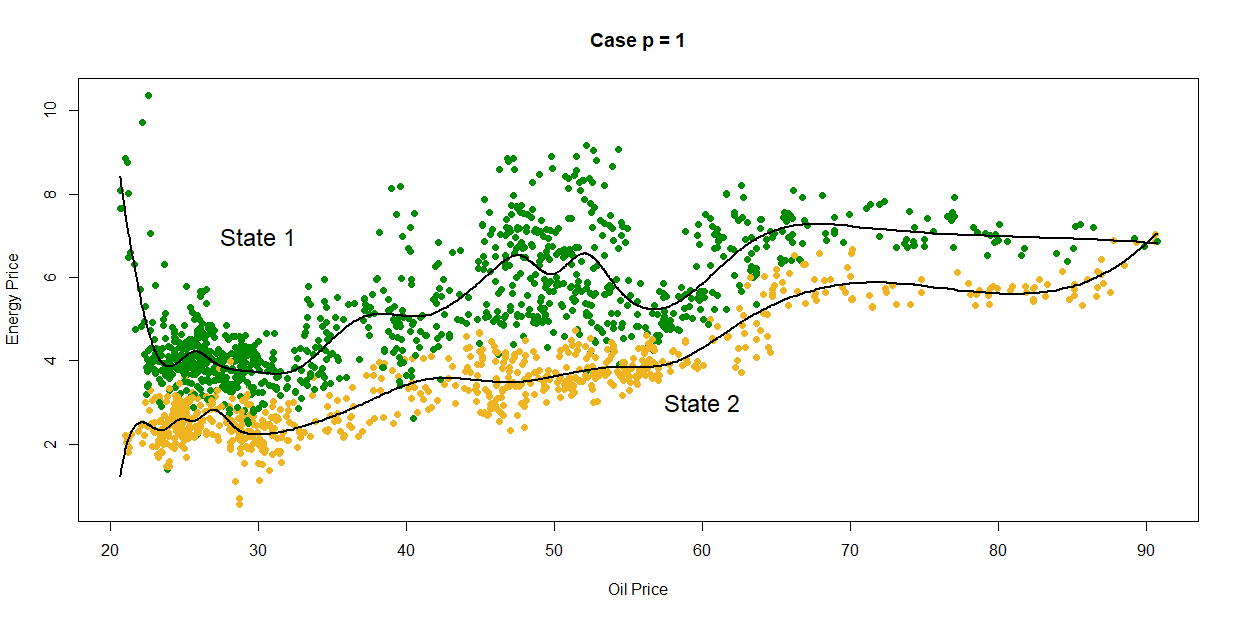}}
\caption{Prediction curves for regime swithching nonparametric regression model of energy price on oil price.}\label{spain3}
\end{figure}

To compare the prediction performance of the above-mentioned model with the simple (single state) additive regression model, we can use the \code{additive\_reg\_mstep} function, with a weight matrix with a single column and all components equal to 1.

\begin{lstlisting}
|\bf \color{lightgray} R$>$| n = energy.hmm$N
|\bf \color{lightgray} R$>$| wt = matrix(rep(1, n), n, 1)
|\bf \color{lightgray} R$>$| emission = additive_reg_mstep(energy.hmm$x, wt,
|\bf \color{lightgray} +|     control = list(K = K))
|\bf \color{lightgray} R$>$| tmpfit = list(model = list(J = 1, parms.emission = emission))
|\bf \color{lightgray} R$>$| pred1 <- addreg_hhsmm_predict(tmpfit, energy.hmm$x[,2:(p+1)], K)
\end{lstlisting}
We compute the sum of squared errors (SSE) of the two competitive models as follows.
\begin{lstlisting}
|\bf \color{lightgray} R$>$| sse1 = sum((pred1 - energy.hmm$x[, 1])^2)
|\bf \color{lightgray} R$>$| sse2 = sum(sapply(1:J, function(j)
|\bf \color{lightgray} +|     sum((pred2[[j]][s == j] - energy.hmm$x[s == j, 1]) ^ 2)))
\end{lstlisting}

We plot the predictions of two competitive models by adding their SSE values to plots, as follows.

\begin{lstlisting}
|\bf \color{lightgray} R$>$| par(mfrow = c(1, 2))
|\bf \color{lightgray} R$>$| plot(energy.hmm$x[, 1], type = "l", xlab = 'Time',
|\bf \color{lightgray} +|     ylab = 'Energy Price', main = "Regime switching
|\bf \color{lightgray} +|     additive regression", lwd = 2)
|\bf \color{lightgray} R$>$| time = 1:length(pred2[[1]])
|\bf \color{lightgray} R$>$| predict = (s == 1) * pred2[[1]] + (s == 2) * pred2[[2]]
|\bf \color{lightgray} R$>$| lines(predict~ time, col = "red", lwd = 2)
|\bf \color{lightgray} R$>$| lines(1:energy.hmm$N, rep(0.5, energy.hmm$N),
|\bf \color{lightgray} +|     col = s, lwd = 2, type = "h", pch = 16)
|\bf \color{lightgray} R$>$| text(500, 9, paste("SSE = ", round(sse2, 2)))
|\bf \color{lightgray} R$>$| plot(energy.hmm$x[, 1], type = "l", xlab = 'Time',
|\bf \color{lightgray} +|     ylab = 'Energy Price', main = "Simple additive
|\bf \color{lightgray} +|     regression", lwd = 2)
|\bf \color{lightgray} R$>$| time = 1:length(pred1)
|\bf \color{lightgray} R$>$| lines(pred1 ~ time, col = "blue", lwd = 2)
|\bf \color{lightgray} R$>$| text(500, 9, paste("SSE = ", round(sse1, 2)))
\end{lstlisting}

\begin{figure}
\centerline{\includegraphics[width = 15cm]{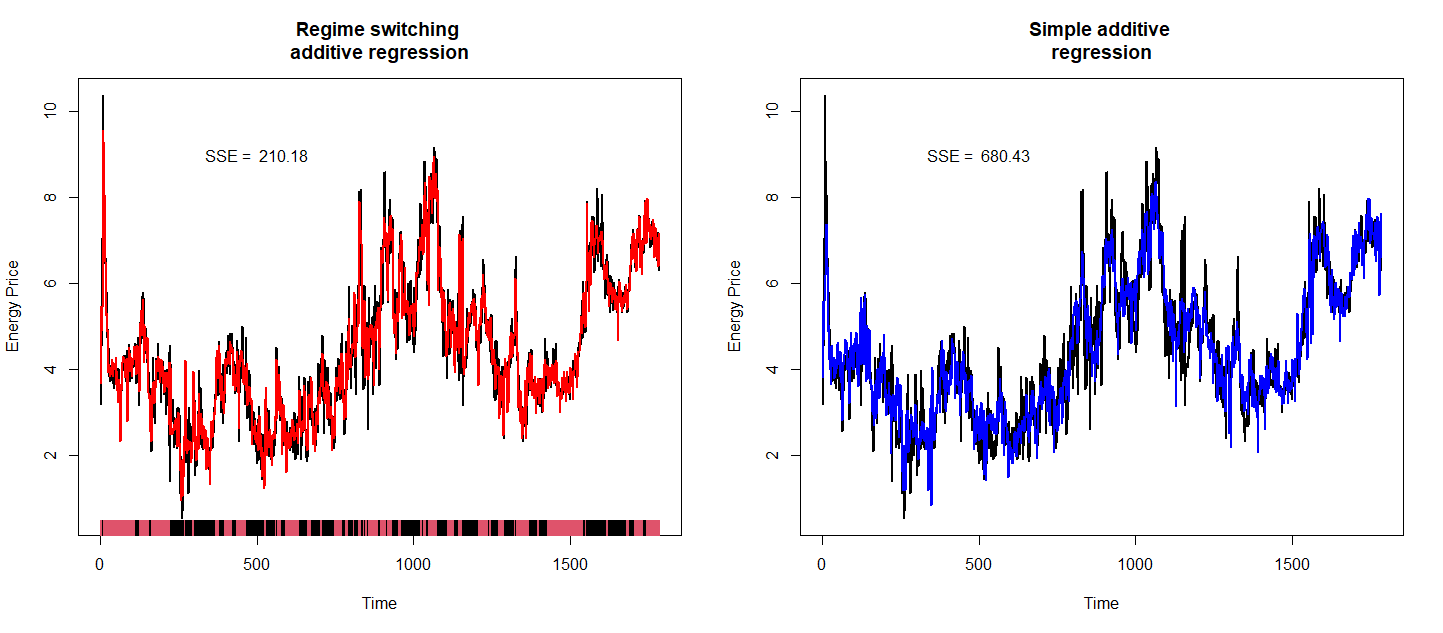}}
\caption{Comparision with simple additive regression.}\label{compare}
\end{figure}

The resulting plot is given in Figure \ref{compare}. As one can see from Figure \ref{compare}, the two-state regime switching additive regression model, performs much better than the simple (single state) additive regression model.
}

\subsection{RUL estimation for the C-MAPSS data set}\label{s6}

The turbofan engine data is from the Prognostic Center of Excellence
(PCoE) of NASA Ames Research Center,
which is simulated by the Commercial Modular Aero-Propulsion
System Simulation (C-MAPSS). Only 14 out of 21 variables are selected, by a method mentioned by \cite{lea15} and are included in the \pkg{hhsmm} package. A list of all 21 variable, as well as a description and the selected 14 variables are tabulated in Table \ref{tabl2}. The \code{train} an \code{test} lists are of class \code{"hhsmmdata"}. The original data set contains the subsets \code{FD001}-\code{FD004}, which are concatenated in the \code{CMAPSS} data set. These sets are described in Table \ref{tabl1}. This table is presented in \code{CMAPSS\$subsets} in the \code{CMAPSS} data set.

\begin{table}[h]
\centering
\caption{C-MAPSS data set overview}\label{tabl1}
\begin{tabular}{c c c c c}
\hline
\hline
 &FD001 & FD002 & FD003 & FD004 \\
\hline
 Training Units & 100 & 260 & 100 & 249\\
 Testing Units & 100 & 259 & 100 & 248\\
 \hline
 \hline
\end{tabular}
\end{table}

\begin{table}[h]
	\centering
\caption{Sensor description of the C-MAPSS data set \citep[see][]{lea15}}	\label{tabl2}
\begin{tabular}{c c c c c}	
\hline
\hline
No. & Symbol & Description & Units & Included in the package? \\
\hline
1 & T2 & Total Temperature at fan inlet &  $^o$R & $\boldsymbol\times$\\
2 & T24 & Total temperature at LPC outlet & $^o$R &\checkmark \\
3 & T30 & Total temperature at HPC outlet  & $^o$R &\checkmark \\
4 & T50 & Total temperature LPT outlet   & $^o$ R&\checkmark \\
5 & P2 & Pressure at fan inlet & psia & $\boldsymbol\times$\\
6 & P15 & Total pressure in bypass-duct & psia & $\boldsymbol\times$\\
7 & P30 & Total pressure at HPC outlet & psia &\checkmark\\
8 & Nf & Physical fan speed & rpm & 	\checkmark \\	
9 & Nc & Physical core speed & rpm & \checkmark \\
10 & Epr & Engine pressure ratio & - & $\boldsymbol\times$\\
11 & Ps30 & Static pressure at HPC outlet & psia & \checkmark\\
12 & Phi & Ratio of fuel flow to Ps30 & pps/psi & \checkmark\\
13 & NRf & Corrected fan speed & rpm & \checkmark\\
14 & NRc & Corrected core speed & rpm & \checkmark\\
15 & BPR & Bypass ratio & $-$ & \checkmark\\
16 & farB & Burner fuel-air ratio & $-$ & $\boldsymbol\times$ \\
17 & htBleed &Bleed enthalpy & $-$ & \checkmark \\
18 & NF dmd & Demanded fan speed & rpm & $\boldsymbol\times$\\
19 & PCNR dmd & Demanded corrected fan speed & rpm & $\boldsymbol\times$\\
20 & W31 & HPT coolant bleed & lbm/s & \checkmark \\
21 & W32 & LPT coolant bleed & lbm/s & \checkmark \\
\hline
\hline
\end{tabular}
\end{table}

\begin{figure}[ht]
\centerline{\includegraphics[width=4.5cm]{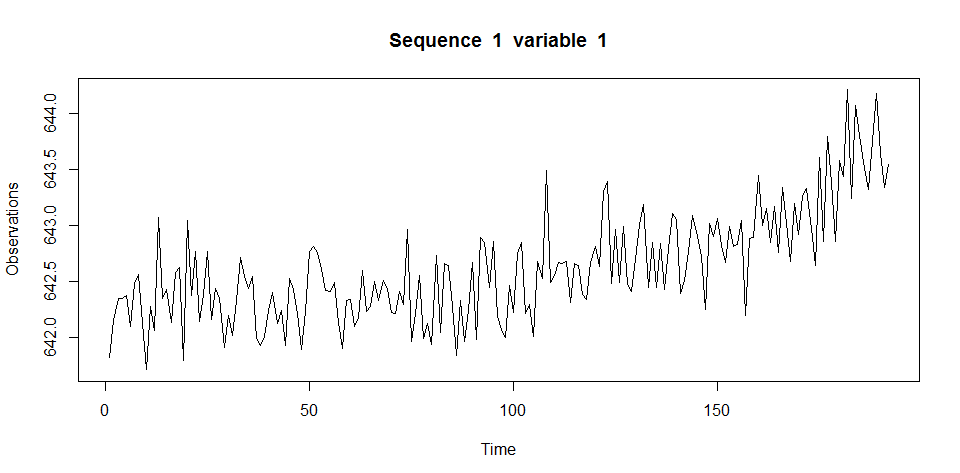}\includegraphics[width=4.5cm]{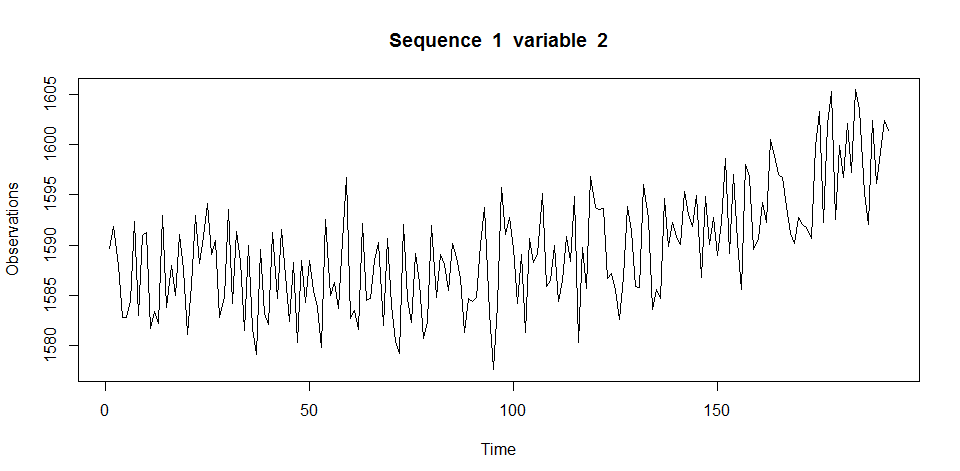}\includegraphics[width=4.5cm]{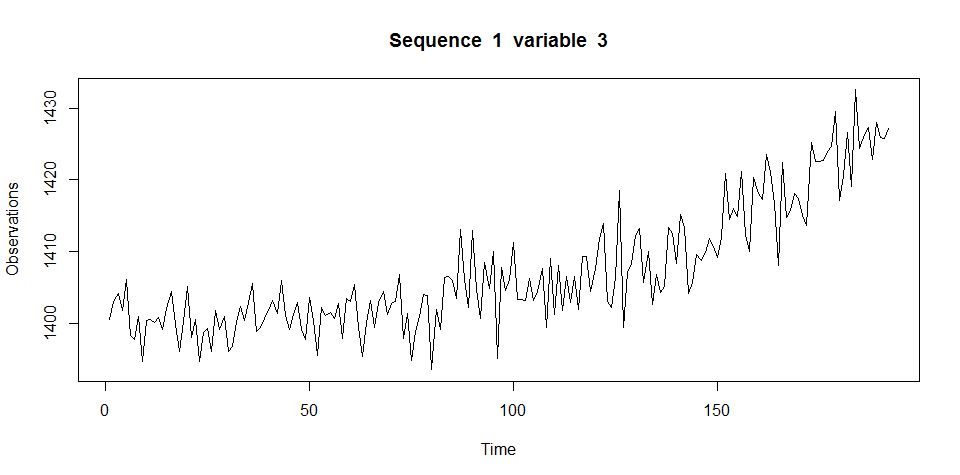}}
\centerline{\includegraphics[width=4.5cm]{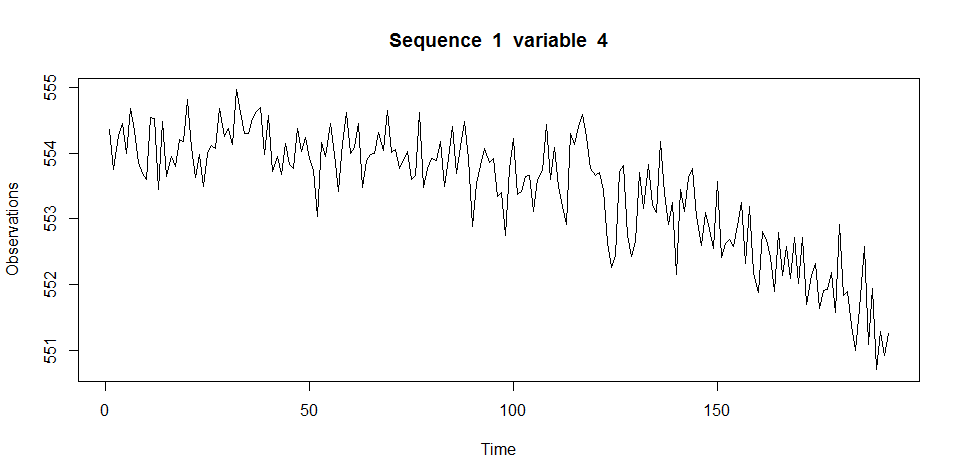}\includegraphics[width=4.5cm]{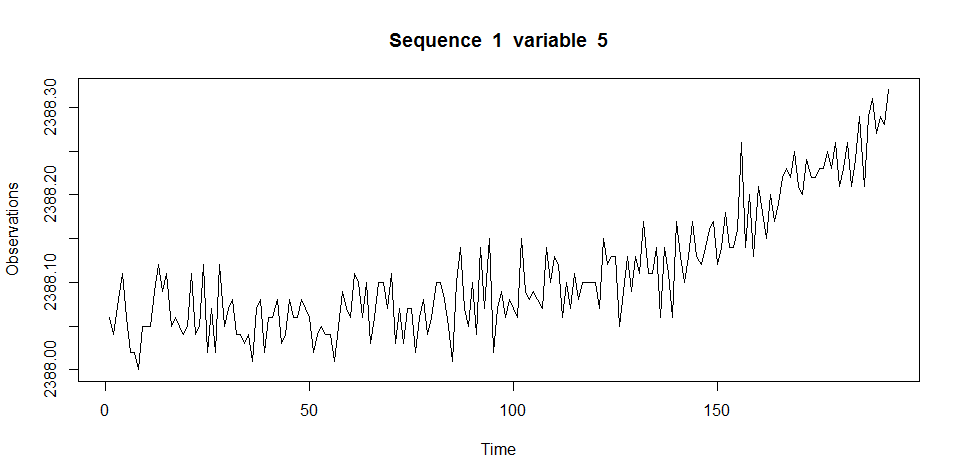}\includegraphics[width=4.5cm]{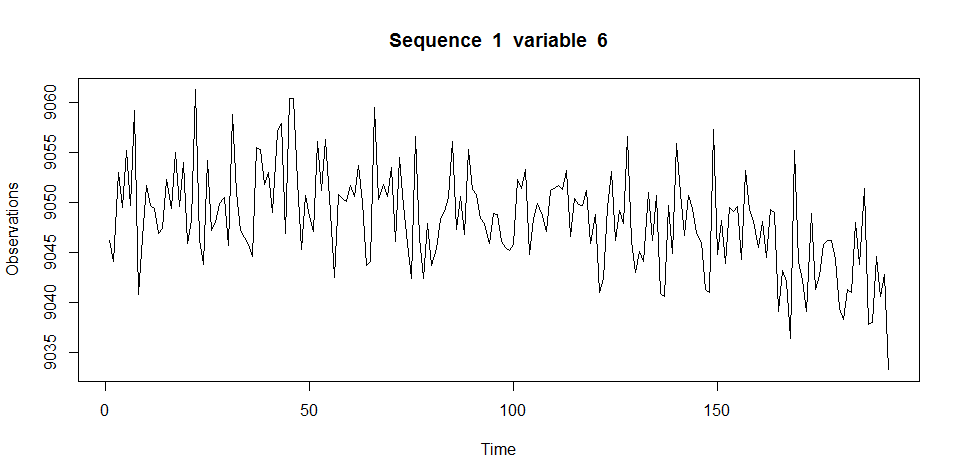}}
\centerline{\includegraphics[width=4.5cm]{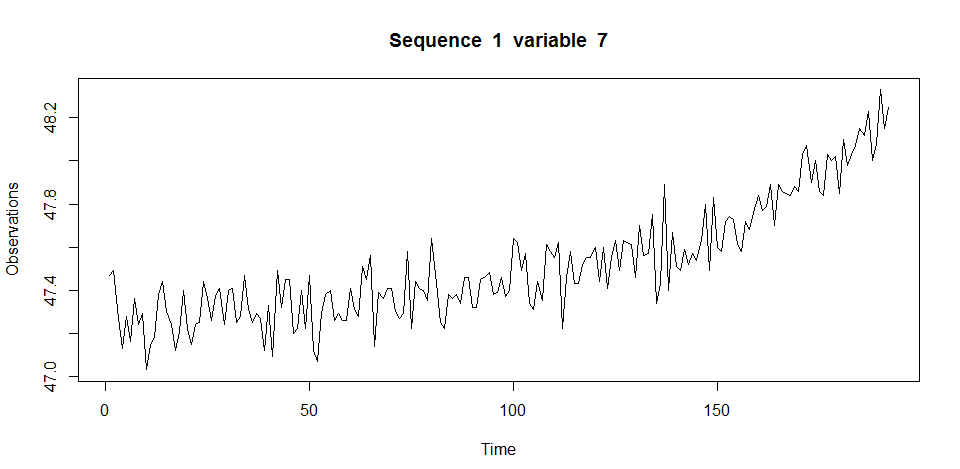}\includegraphics[width=4.5cm]{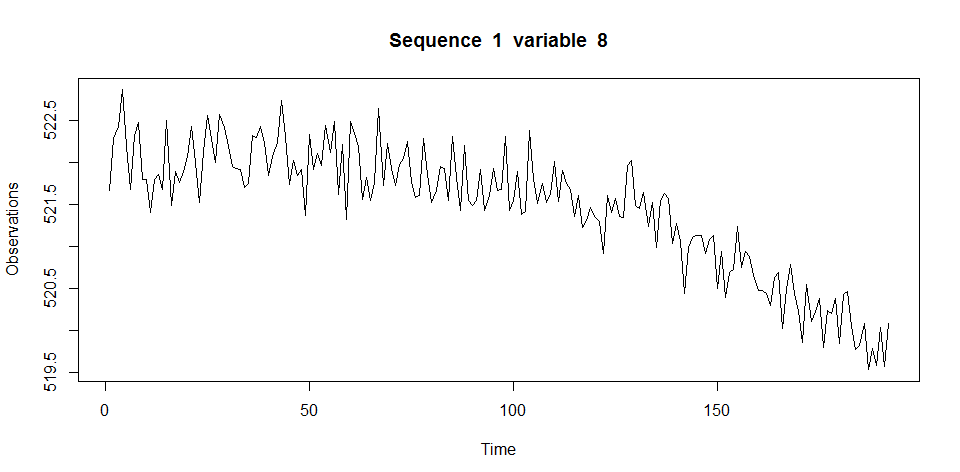}\includegraphics[width=4.5cm]{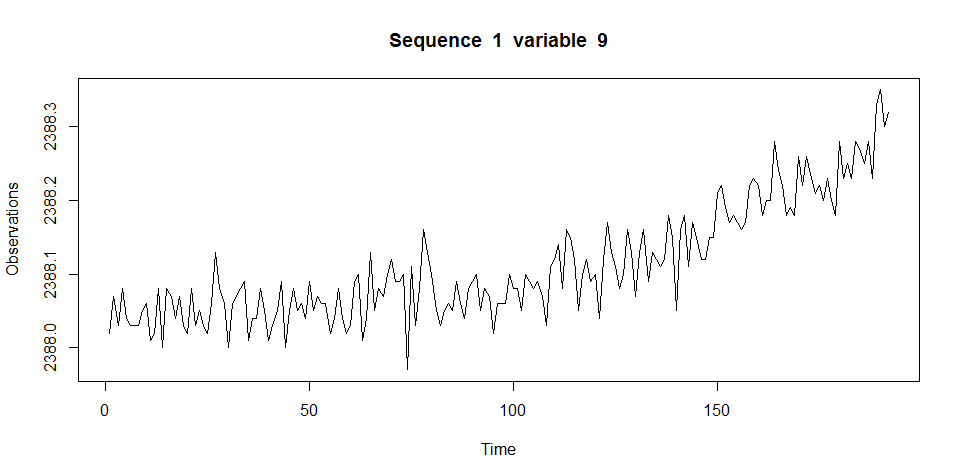}}
\centerline{\includegraphics[width=4.5cm]{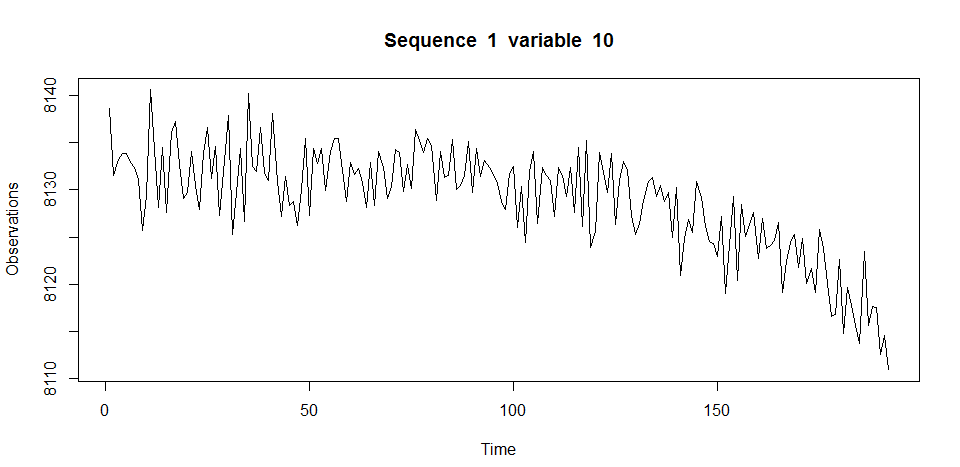}\includegraphics[width=4.5cm]{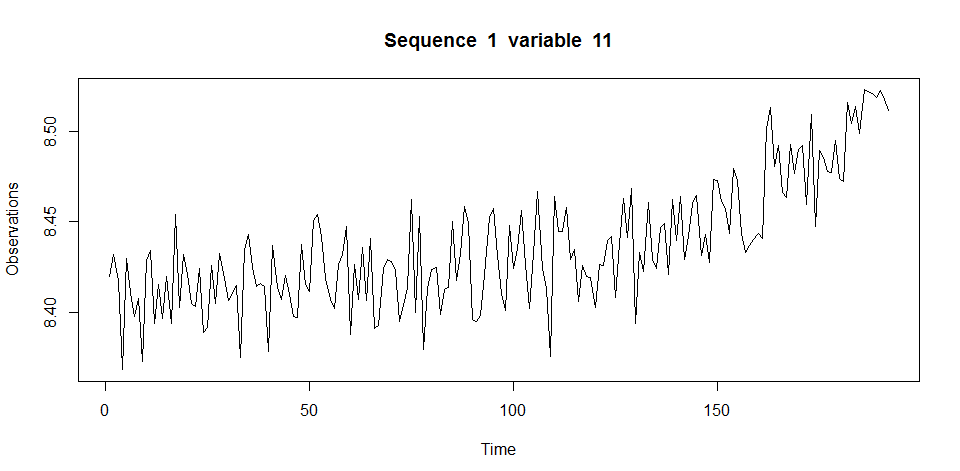}\includegraphics[width=4.5cm]{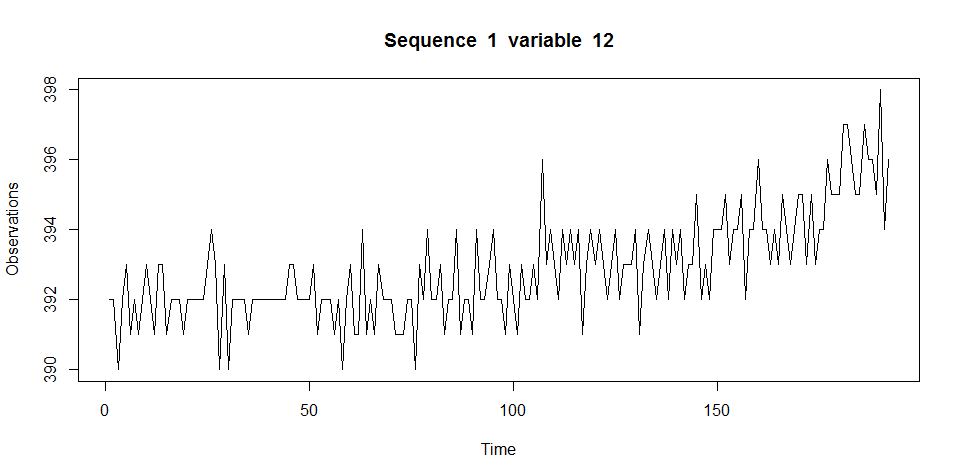}}
\centerline{\includegraphics[width=4.5cm]{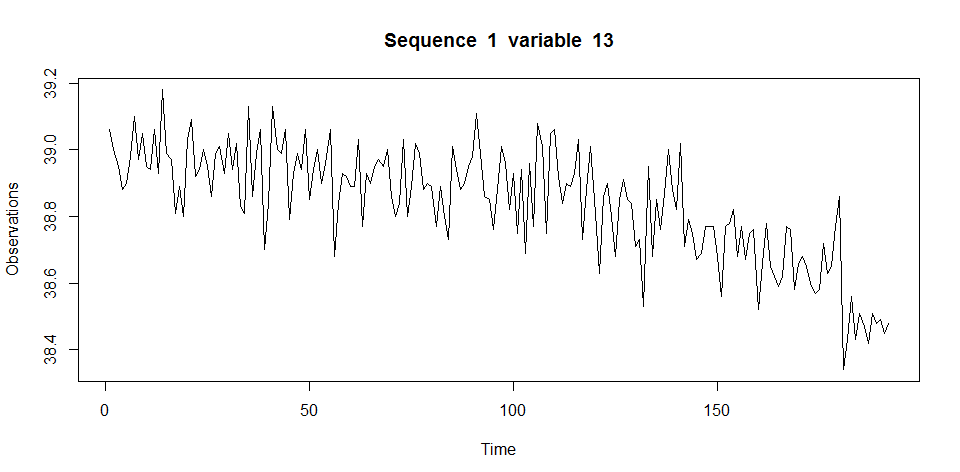}\includegraphics[width=4.5cm]{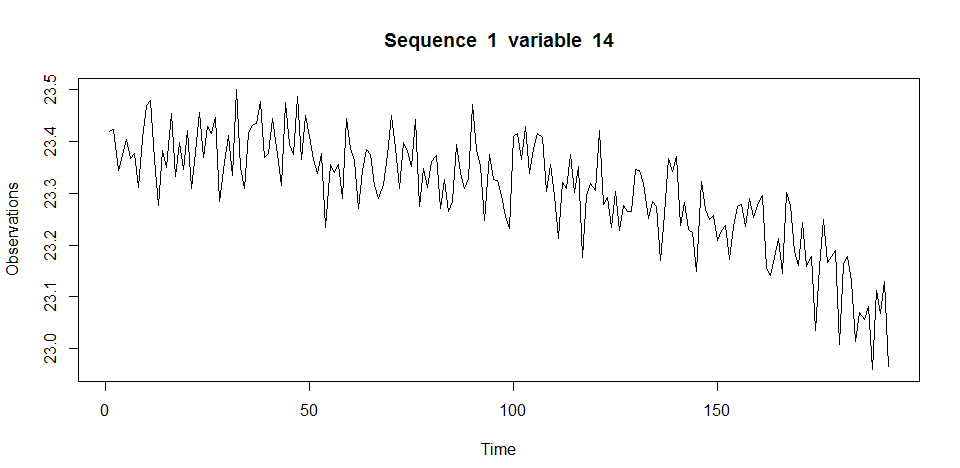}}
\caption{The time series plot of 14 variables of the first sequence of the \code{train} set for the \code{CMAPSS} data set. }\label{tsp}
\end{figure}

We load the data set and extract the \code{train} and \code{test} sets as follows.
\begin{lstlisting}
|\bf \color{lightgray} R$>$| data(CMAPSS)
|\bf \color{lightgray} R$>$| train = CMAPSS$train
|\bf \color{lightgray} R$>$| test = CMAPSS$test
\end{lstlisting}
To visualize the data set, we plot only the first sequence of the \code{train} set. To do this, this sequence is converted to a data set of class \code{"hhsmmdata"}, using the function \code{hhsmmdata} as follows. The plots are presented in Figure \ref{tsp}.
\begin{lstlisting}
|\bf \color{lightgray} R$>$| train1 = hhsmmdata(x = train$x[1:train$N[1],], N = train$N[1])
|\bf \color{lightgray} R$>$| plot(train1)
\end{lstlisting}

Initial clustering of the states and mixture components is obtained by the \code{initial\_cluster} function. Since, the suitable reliability model for the \code{CMAPSS} data set is a left to right model, the option \code{ltr = TRUE} is used. Also, since the engines are failed in the final time of each sequence, the final time of each sequence is considered the absorbing state (final state of the left to right model). This assumption is given to the model by the option \code{final.absorb = TRUE}. The number of states is assumed to be 5 states, which could be one healthy state, 3 levels of damage state, and one failure state in the reliability model. The number of mixture components is computed automatically using the option \code{nmix = "auto"}.
\begin{lstlisting}
|\bf \color{lightgray} R$>$| J = 5
|\bf \color{lightgray} R$>$| clus = initial_cluster(train = train, nstate = J, nmix = "auto",
|\bf \color{lightgray} +|     ltr = TRUE, final.absorb = TRUE, verbose = TRUE)
Within sequence clustering ...
clustering [=========================]   100% in  3m
State  1
Between sequence clustering ...
Automatic determination of the number of mixture components ...
State  2
Between sequence clustering ...
Automatic determination of the number of mixture components ...
State  3
Between sequence clustering ...
Automatic determination of the number of mixture components ...
State  4
Between sequence clustering ...
Automatic determination of the number of mixture components ...
State  5
Between sequence clustering ...
Automatic determination of the number of mixture components ...
\end{lstlisting}
Now, we initialize the model using the \code{initialize\_model} function. The sojourn time distribution is assumed to be \code{"gamma"} distribution.
\begin{lstlisting}
|\bf \color{lightgray} R$>$| initmodel = initialize_model(clus = clus, sojourn = "gamma",
|\bf \color{lightgray} +|     M = max(train$N), verbose = TRUE)
Intitial estimation ....
State  1  estimation
Mixture component  1  estimation
Mixture component  2  estimation
Mixture component  3  estimation
Mixture component  4  estimation
Mixture component  5  estimation
...
State  5  estimation
Mixture component  1  estimation
Mixture component  2  estimation
Mixture component  3  estimation
Mixture component  4  estimation
Mixture component  5  estimation
Initializing model ...
\end{lstlisting}

\begin{figure}
\centerline{\includegraphics[width = 15cm]{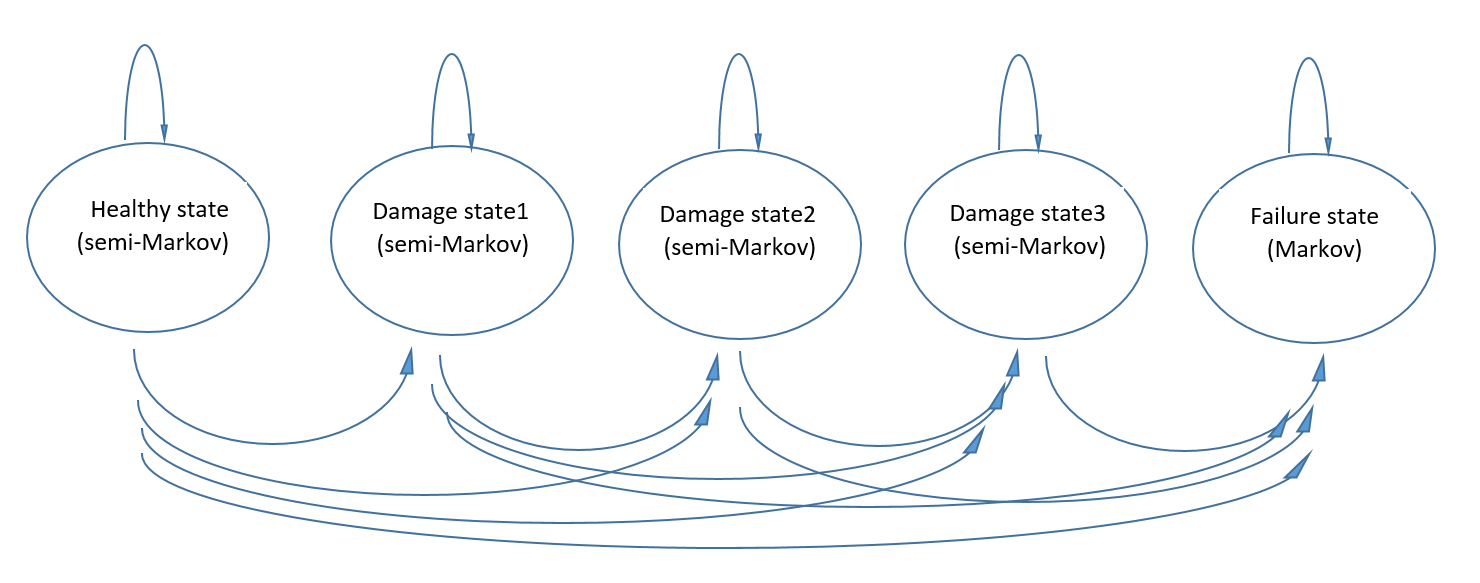}}
\caption{Graphical representation of the reliability left to right model.}\label{rtl}
\end{figure}

\begin{figure}
\centerline{\includegraphics[width=10cm]{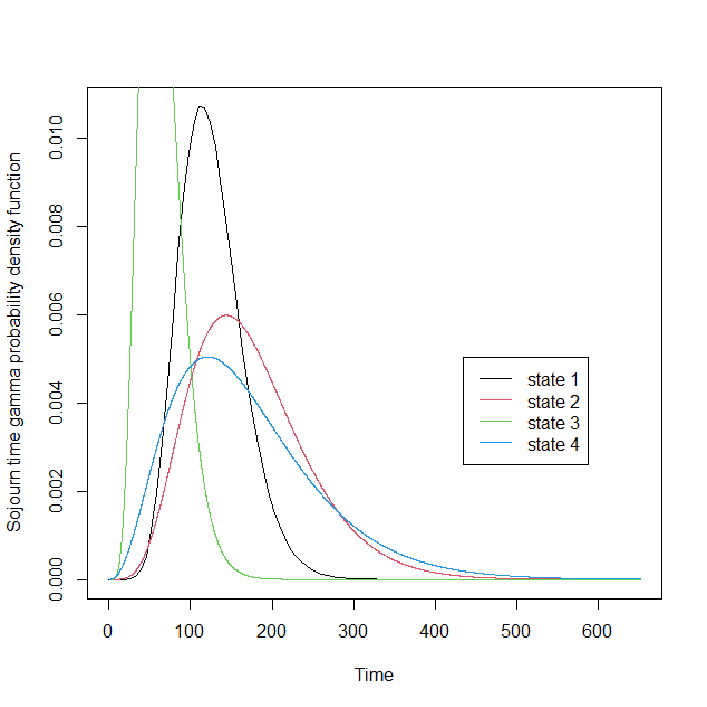}}
\caption{The estimated gamma sojourn time density functions.}\label{gamfit}
\end{figure}

As a result, the initial estimates of the parameters of the sojourn time distribution and initial estimates of the transition probability matrix and the initial probability vector are obtained as follows.
\begin{lstlisting}
|\bf \color{lightgray} R$>$| initmodel$sojourn
$shape
[1] 10.753944 11.222101  5.617826  8.386559  0.000000
$scale
[1]  0.9138905  0.8885851  1.9869251 23.1578163  0.0000000
$type
[1] "gamma"
|\bf \color{lightgray} R$>$| initmodel$transition
     [,1] [,2]      [,3]       [,4]       [,5]
[1,]    0 0.85 0.0500000 0.05000000 0.05000000
[2,]    0 0.00 0.8947368 0.05263158 0.05263158
[3,]    0 0.00 0.0000000 0.94444444 0.05555556
[4,]    0 0.00 0.0000000 0.00000000 1.00000000
[5,]    0 0.00 0.0000000 0.00000000 1.00000000
|\bf \color{lightgray} R$>$| initmodel$init
[1] 1 0 0 0 0
\end{lstlisting}
Now, we fit the HHSMM model, using \code{hhsmmfit} function. The option \code{lock.init=TRUE} is a good option for a right to left model since the initial state is the first state (healthy system state in the reliability model) in such situations with probability 1. Graphical visualization of such a model is given in Figure \ref{rtl}.
\begin{lstlisting}
|\bf \color{lightgray} R$>$| fit1 = hhsmmfit(x = train, model = initmodel, M = max(train$N),
|\bf \color{lightgray} +|     par = list(lock.init = TRUE))
iteration:  1   log-likelihood =  -1813700
iteration:  2   log-likelihood =  -1381056
iteration:  3   log-likelihood =  -1470325
iteration:  4   log-likelihood =  -1471163
iteration:  5   log-likelihood =  -1450113
iteration:  6   log-likelihood =  -1429393
....
iteration:  79   log-likelihood =  -1380078
iteration:  80   log-likelihood =  -1379937
iteration:  81   log-likelihood =  -1379810
AIC =  2770213
BIC =  2823105
\end{lstlisting}
The estimates of the transition probability matrix, the sojourn time probability matrix, the initial probability vector, and the AIC and BIC of the model, are extracted as follows.
\begin{lstlisting}
|\bf \color{lightgray} R$>$| fit1$model$transition
     [,1]      [,2]      [,3]       [,4] [,5]
[1,]    0 0.1481077 0.7954866 0.05640566    0
[2,]    0 0.0000000 0.0000000 1.00000000    0
[3,]    0 0.0000000 0.0000000 0.00000000    1
[4,]    0 0.0000000 0.0000000 0.00000000    1
[5,]    0 0.0000000 0.0000000 0.00000000    1
|\bf \color{lightgray} R$>$| head(fit1$model$d)
             [,1]         [,2]         [,3]         [,4]   [,5]
[1,] 3.302753e-19 5.011609e-12 2.064632e-10 1.101749e-07 1e-100
[2,] 4.526557e-16 2.694189e-10 1.604557e-08 1.111672e-06 1e-100
[3,] 2.957092e-14 2.545406e-09 1.854185e-07 3.737021e-06 1e-100
[4,] 5.455948e-13 1.178541e-08 9.730204e-07 8.376700e-06 1e-100
[5,] 5.017141e-12 3.736808e-08 3.346571e-06 1.528679e-05 1e-100
[6,] 2.975970e-11 9.388985e-08 8.869356e-06 2.464430e-05 1e-100
|\bf \color{lightgray} R$>$| fit1$model$init
[1] 1 0 0 0 0
|\bf \color{lightgray} R$>$| fit1$AIC
[1] 2770213
|\bf \color{lightgray} R$>$| fit1$BIC
[1] 2823105
\end{lstlisting}
We can plot the estimated gamma sojourn probability density functions as follows.
\begin{lstlisting}
|\bf \color{lightgray} R$>$| J <- 5
|\bf \color{lightgray} R$>$| MM = max(train$N) * 1.2
|\bf \color{lightgray} R$>$| f1 <- function(x) dgamma(x, shape = fit1$model$sojourn$shape[1],
|\bf \color{lightgray} +|     scale = fit1$model$sojourn$scale[1])
|\bf \color{lightgray} R$>$| plot(f1, 0, MM, type = "l", xlab = "Time", ylab =
|\bf \color{lightgray} +|     "Sojourn time gamma probability density function")
|\bf \color{lightgray} R$>$| leg = "state 1"
|\bf \color{lightgray} R$>$| for(j in 2:(J-1)){
|\bf \color{lightgray} +|     f <- function(x) dgamma(x, shape = fit1$model$sojourn$shape[j],
|\bf \color{lightgray} +|     scale = fit1$model$sojourn$scale[j])
|\bf \color{lightgray} +|     xs <- seq(1, MM, 0.1)
|\bf \color{lightgray} +|     ys <- sapply(xs, f)
|\bf \color{lightgray} +|     lines(xs, ys, col = j)
|\bf \color{lightgray} +|     leg = c(leg, paste("state", j))
|\bf \color{lightgray} +|    }
|\bf \color{lightgray} R$>$| legend(2*MM/3, max(ys), leg, lty = rep(1, J - 1), col = 1:(J - 1))
\end{lstlisting}
The resulted plot is shown in Figure \ref{gamfit}.

Now, we obtain the estimates of the RULs, as well as the confidence intervals, by four different methods as follows. These four methods are obtained by combination of two different methods \code{"viterbi"} and \code{"smoothing"} for the prediction and two different methods \code{"mean"} and \code{"max"} for RUL estimation and confidence interval computation. The option \code{"viterbi"} uses the Viterbi algorithm to find the most likely state sequence, while the option \code{"smoothing"} uses the estimates of the state probabilities, using the emission probabilities of the \code{test} data set. On the other hand, in the \code{"mean"} method, the mean sojourn time and its standard deviation are used for estimation and confidence interval, while in the \code{"max"} method, the maximum probability sojourn time and its quantiles are used (see Section \ref{s4}).

\begin{lstlisting}
|\bf \color{lightgray} R$>$| pp1 = predict(fit1, test, method = "viterbi",
|\bf \color{lightgray} +|     RUL.estimate = TRUE, confidence = "mean")
|\bf \color{lightgray} R$>$| pp2 = predict(fit1, test, method = "viterbi",
|\bf \color{lightgray} +|     RUL.estimate = TRUE, confidence = "max")
|\bf \color{lightgray} R$>$| pp3 = predict(fit1, test, method = "smoothing",
|\bf \color{lightgray} +|     RUL.estimate = TRUE, confidence = "mean")
|\bf \color{lightgray} R$>$| pp4 = predict(fit1, test, method = "smoothing",
|\bf \color{lightgray} +|     RUL.estimate = TRUE, confidence = "max")
\end{lstlisting}
As a competitor, we fit the hidden Markov model (HMM) to the data set, which means that we consider all states to be Markovian. To de this,
we try fitting the HMM to the \code{train} set, using the option \code{semi = rep(FALSE,J)} of the \code{hhsmmfit} function of the \pkg{hhsmm} package. We use the same initial values of the parameters, while we need to use \code{dens.emission = dmixmvnorm} in the \code{hhsmmspec} function, and set the mixture components probabilities equal to 1 (for one mixture component in each state).
\begin{lstlisting}
|\bf \color{lightgray} R$>$| J <- 5
|\bf \color{lightgray} R$>$| init0 <- c(1, 0, 0, 0, 0)
|\bf \color{lightgray} R$>$| P0 <- initmodel$transition
|\bf \color{lightgray} R$>$| b0 = list(mu = list(), sigma = list())
|\bf \color{lightgray} R$>$| for(j in 1:J){
|\bf \color{lightgray} +|     b0$mu[[j]] <- Reduce('+', initmodel$parms.emission$mu[[j]]) / J
|\bf \color{lightgray} +|     b0$sigma[[j]] <- Reduce('+',
|\bf \color{lightgray} +|         initmodel$parms.emission$sigma[[j]]) / J
|\bf \color{lightgray} +|  }
|\bf \color{lightgray} R$>$| for(j in 1:J) b0$mix.p[[j]] = 1
|\bf \color{lightgray} R$>$| initmodel <- hhsmmspec(init = init0, transition = P0,
|\bf \color{lightgray} +|     parms.emission = b0, dens.emission = dmixmvnorm,
|\bf \color{lightgray} +|     semi = rep(FALSE, J))
|\bf \color{lightgray} R$>$| fit3 = hhsmmfit(train, initmodel , mstep = mixmvnorm_mstep)
iteration:  1   log-likelihood =  -110772073
iteration:  2   log-likelihood =  -1228418
iteration:  3   log-likelihood =  -1246009
iteration:  4   log-likelihood =  -1260494
iteration:  5   log-likelihood =  -1284620
iteration:  6   log-likelihood =  -1329087
iteration:  7   log-likelihood =  -1506447
iteration:  8   log-likelihood =  -2650675
iteration:  9   log-likelihood =  -2652023
iteration:  10   log-likelihood =  -2650571
iteration:  11   log-likelihood =  -2650527
AIC =  5303215
BIC =  5313999
\end{lstlisting}
For the fitted HMM model, we estimate the RULs using the aforementioned options.
\begin{lstlisting}
|\bf \color{lightgray} R$>$| pp5 = predict(fit3, test, method = "viterbi",
|\bf \color{lightgray} +|     RUL.estimate = TRUE, confidence = "mean")
|\bf \color{lightgray} R$>$| pp6 = predict(fit3, test, method = "viterbi",
|\bf \color{lightgray} +|     RUL.estimate = TRUE, confidence = "max")
|\bf \color{lightgray} R$>$| pp7 = predict(fit3, test, method = "smoothing",
|\bf \color{lightgray} +|     RUL.estimate = TRUE, confidence = "mean")
|\bf \color{lightgray} R$>$| pp8 = predict(fit3, test, method = "smoothing",
|\bf \color{lightgray} +|     RUL.estimate = TRUE, confidence = "max")
\end{lstlisting}
Now, we use the real values of the RULs, stored in \code{test\$RUL} to compute the coverage probabilities of the confidence intervals of HHSMM and HMM models.
\begin{lstlisting}
|\bf \color{lightgray} R$>$| mean((test$RUL >= pp1$RUL.low) & (test$RUL <= pp1$RUL.up))
[1] 0.7963225
|\bf \color{lightgray} R$>$| mean((test$RUL >= pp2$RUL.low) & (test$RUL <= pp2$RUL.up))
[1] 0.54314
|\bf \color{lightgray} R$>$| mean((test$RUL >= pp3$RUL.low) & (test$RUL <= pp3$RUL.up))
[1] 0.864215
|\bf \color{lightgray} R$>$| mean((test$RUL >= pp4$RUL.low) & (test$RUL <= pp4$RUL.up))
[1] 0.7213579
|\bf \color{lightgray} R$>$| mean((test$RUL >= pp5$RUL.low) & (test$RUL <= pp5$RUL.up))
[1] 0
|\bf \color{lightgray} R$>$| mean((test$RUL >= pp6$RUL.low) & (test$RUL <= pp6$RUL.up))
[1] 0
|\bf \color{lightgray} R$>$| mean((test$RUL >= pp7$RUL.low) & (test$RUL <= pp7$RUL.up))
[1] 0
|\bf \color{lightgray} R$>$| mean((test$RUL >= pp8$RUL.low) & (test$RUL <= pp8$RUL.up))
[1] 0
\end{lstlisting}
As one can see from the above results, the HHSMM model's coverage probabilities are much better than the HMM ones.

\begin{figure}
\centerline{\includegraphics[width = 15cm]{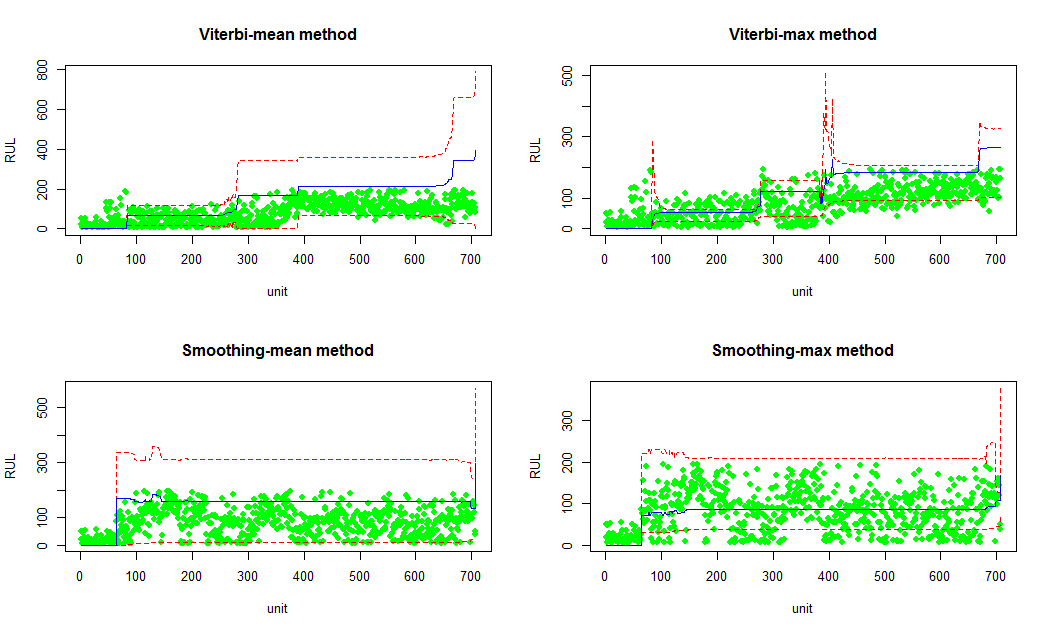}}
\caption{RUL estimates (solid blue lines) and RUL bounds (dashed red lines) using four different methods for the CMAPSS test data set.}\label{RUL}
\end{figure}

To visualize the results of RUL estimation, we plot the RUL estimates and RUL bounds as follows.
\begin{lstlisting}
|\bf \color{lightgray} R$>$| par(mfrow = c(2,2))
|\bf \color{lightgray} R$>$| plot(test$RUL[order(pp1$RUL)],
|\bf \color{lightgray} +|     ylim = c(min(pp1$RUL.low), max(pp1$RUL.up)),
|\bf \color{lightgray} +|     pch=16, col = "green", xlab = "unit",
|\bf \color{lightgray} +|     ylab = "RUL", main = "Viterbi-mean method")
|\bf \color{lightgray} R$>$| lines(pp1$RUL.low[order(pp1$RUL)], lty = 2, col = "red")
|\bf \color{lightgray} R$>$| lines(pp1$RUL.up[order(pp1$RUL)], lty = 2,
|\bf \color{lightgray} +|     col = "red")
|\bf \color{lightgray} R$>$| lines(pp1$RUL[order(pp1$RUL)], col = "blue")
|\bf \color{lightgray} R$>$| plot(test$RUL[order(pp2$RUL.low)],
|\bf \color{lightgray} +|     ylim = c(min(pp2$RUL.low), max(pp2$RUL.up)),
|\bf \color{lightgray} +|     pch=16, col = "green", xlab = "unit",
|\bf \color{lightgray} +|     ylab = "RUL", main = "Viterbi-max method")
|\bf \color{lightgray} R$>$| lines(sort(pp2$RUL.low), lty = 2, col = "red")
|\bf \color{lightgray} R$>$| lines(pp2$RUL.up[order(pp2$RUL.low)], lty = 2,
|\bf \color{lightgray} +|     col = "red")
|\bf \color{lightgray} R$>$| lines(pp2$RUL[order(pp2$RUL.low)], col = "blue")
|\bf \color{lightgray} R$>$| plot(test$RUL[order(pp3$RUL.low)],
|\bf \color{lightgray} +|     ylim = c(min(pp3$RUL.low), max(pp3$RUL.up)),
|\bf \color{lightgray} +|     pch=16, col = "green", xlab = "unit",
|\bf \color{lightgray} +|     ylab = "RUL", main = "Smoothing-mean method")
|\bf \color{lightgray} R$>$| lines(sort(pp3$RUL.low), lty = 2, col = "red")
|\bf \color{lightgray} R$>$| lines(pp3$RUL.up[order(pp3$RUL.low)], lty = 2,
|\bf \color{lightgray} +|     col = "red")
|\bf \color{lightgray} R$>$| lines(pp3$RUL[order(pp3$RUL.low)], col = "blue")
|\bf \color{lightgray} R$>$| plot(test$RUL[order(pp4$RUL.low)],
|\bf \color{lightgray} +|     ylim = c(min(pp4$RUL.low), max(pp4$RUL.up)),
|\bf \color{lightgray} +|     pch=16, col = "green", xlab = "unit",
|\bf \color{lightgray} +|     ylab = "RUL", main = "Smoothing-max method")
|\bf \color{lightgray} R$>$| lines(sort(pp4$RUL.low), lty = 2, col = "red")
|\bf \color{lightgray} R$>$| lines(pp4$RUL.up[order(pp4$RUL.low)], lty = 2,
|\bf \color{lightgray} +|     col = "red")
|\bf \color{lightgray} R$>$| lines(pp4$RUL[order(pp4$RUL.low)], col = "blue")
\end{lstlisting}
The resulting plots are presented in Figure \ref{RUL}. From Figure \ref{RUL} and the above coverage probabilities, one can see that the ``smoothing" and ``max" methods perform better than other methods, in this example.

{\section{Concluding remarks}

This paper presents several examples of the \proglang{R} package \pkg{hhsmm}. The scope of application of this package covers simulation, initialization, fitting, and prediction of HMM, HSMM, and HHSMM models, for different types of discrete and continuous sojourn distribution, including shifted Poisson, negative binomial, logarithmic, gamma, Weibull, and log-normal. This package contains density and M-step function for estimation of the emission distribution for different types of emission distribution, including the mixture of multivariate normals and penalized B-spline estimator of the emission distribution, the mixture of linear and additive regression (conditional multivariate normal distributions of the response given the covariates; regime-switching regression models) as well as the ability to define another emission distributions by the user. As a special case of the regime-switching regression models, the auto-regressive HHSMM models can be modeled by the \pkg{hhsmm} package. The left to right models are considered in the \pkg{hhsmm} package, especially in the initialization functions. The \pkg{hhsmm} package uses the EM algorithm to handle the missing values when the mixture of multivariate normals is considered as the emission distribution. The ability to predict the future states, residual useful lifetime estimation for the left to right models, computation of the score of new observations, computing the homogeneity of two sequences of states, and splitting the data to train and test sequences by the ability to right-trim the test sequences, are other useful features of the \pkg{hhsmm} package. The current version 0.3.2 of this package is now available on CRAN (\url{https://cran.r-project.org/package=hhsmm}), while the future improvements of this package are also considered by the authors. Any report of the possible bugs of the \pkg{hhsmm} package are welcome through \url{https://github.com/mortamini/hhsmm/issues} and we welcome the users' offers for any needed feature of the package in the future.

\section*{Acknowledgements}

The authors would like to thank the two anonymous referees and the associate editor for their useful comments and suggestions, which improved an earlier version of the \pkg{hhsmm} package and this paper. }

\appendix

\section*{Appendix}
\renewcommand{\theequation}{A\thesection.\arabic{equation}}

\subsection*{Forward-backward algorithm for the HHSMM model}
Denote the sequence $\{Y_{t},\ldots,Y_{s}\}$ by $Y_{t:s}$ and suppose that the observation sequence $X_{0:\tau-1}$ with the corresponding hidden state sequence $S_{0:\tau-1}$ is observed.
The forward-backward algorithm is an algorithm to compute the probabilities
$$L_j(t)=P(S_t=j|X_{0:{\tau-1}}=x_{0:{\tau-1}})$$
within the E-step of the EM algorithm. The above probabilities are computed in the backward recursion of the forward-backward algorithm.

For a semi-Markovian state $j$, and $t=0,\ldots,\tau-2$, the forward recursion computes the following probabilities \citep{g05},

\begin{eqnarray}\label{10eq}
F_j(t) &=& P(S_{t+1}\neq j,S_t = j|X_{0:t} = x_{0:t})\nonumber\\
&=&\frac{f_j(x_t)}{N_t}\Biggl[\sum_{{u=1}}^{t}\Bigg\{\prod_{\nu=1}^{u-1}\frac{f_j(x_{t-\nu})}{N_{t-\nu}}\Bigg\}d_j(u)\sum_{i\neq j}p_{ij}F_i(t-u)\nonumber\\
\qquad &&+\Bigg\{\prod_{\nu=1}^{t}\frac{f_j(x_{t-\nu})}{N_{t-\nu}}\Bigg\}d_j(t+1)\pi_j\Biggl],
\end{eqnarray}
and
 \begin{eqnarray}\label{11eq}
 F_j(\tau-1)&=&P(S_{\tau-1} = j|X_{0:{\tau-1}} =x_{0:{\tau-1}})\nonumber\\
&=&\frac{f_j(x_{\tau-1})}{N_{\tau-1}}\Biggl[\sum_{u=1}^{\tau-1}\Bigg\{\prod_{\nu=1}^{u-1}\frac{f_j(x_{\tau-1-\nu})}{N_{\tau-1-\nu}}\Bigg\}D_j(u)\sum_{i\neq j}p_{ij}F_i(\tau-1-u)\nonumber\\
\qquad &&+\Bigg\{\prod_{\nu=1}^{\tau-1}\frac{f_j(x_{\tau-1-\nu})}{N_{\tau-1-\nu}}\Bigg\}D_j(\tau)\pi_j\Biggl].
 \end{eqnarray}
where the normalizing factor $N_t$ is computed as follows
$$
N_t = P(X_t = x_t|X_{0:{t-1}} = x_{0:{t-1}}) =\sum_{j}P(S_t = j,X_t = x_t|X_{0:{t-1}} = x_{0:{t-1}})
$$
For a Markovian state $j$, and for $t = 0$,
$$ \tilde{F}_j(0) = P(S_0 = j|X_0 = x_0)=\frac{f_j(x_0)}{N_0}\pi_j$$
and for   $t = 1,\ldots,\tau-1$
\begin{equation}\label{7_eq}
\tilde{F}_j(t) = P(S_t = j|X_{0:t} = x_{0:t}) =\frac{f_j(x_t)}{N_t}\sum_{i}\tilde{p}_{ij}\tilde{F}_i(t-1)
  \end{equation}

The log-likelihood of the model is then
$$\log P(X_{0:{\tau-1}} = x_{0:{\tau-1}};\theta) = \sum_{t=0}^{\tau-1}\log N_t,$$
which is used as a criteria for convergence of the EM algorithm and the evaluate the quality of the model.
The backward recursion is initialized by
$$L_j(\tau-1) = P(S_{\tau-1} = j|X_{0:{\tau-1}} = x_{0:{\tau-1}}) = F_j(\tau-1) = \tilde{F}_j(\tau-1).$$
For a semi-Markovian state $j$, we have
$$L_j(\tau-1) = P(S_{\tau-1} = j|X_{0:{\tau-1}} = x_{0:{\tau-1}}) = F_j(\tau-1),
$$
and for $t = \tau-2,\ldots,0$,
\begin{eqnarray}\label{13eq}
L_j(t)&=& L1_j(t) + L_j(t+1) - G_j(t+1)\sum_{i\neq j}{p}_{ij}\tilde{F}_i(t),
\end{eqnarray}
where
\begin{eqnarray}\label{14eq}
L1_j(t)&=& \sum_{k\neq j}G_k(t+1)p_{jk}F_j(t),
\end{eqnarray}
and
\begin{eqnarray*}
G_j(t+1) &=& \sum_{u=1}^{\tau-1-t}G_j(t+1,u)
\end{eqnarray*}
with
$$
G_j(t+1,u)=\left\{\prod_{\nu=0}^{\tau-2-t}\frac{f_j(x_{\tau-1-\nu})}{N_{\tau-1-\nu}}\right\}D_j(\tau-1-t).
$$
For a Markovian state $j$ and for $t = \tau -2,\ldots,0$
\begin{eqnarray}\label{13_eq}
L_j(t)&=&L1_j(t)
\end{eqnarray}
Furthermore, if a mixture of multivariate normal distributions with probability density function
\begin{equation}\label{mmn}
f_j(x) = \sum_{k=1}^{K_j}\lambda_{kj} {\cal{N}}_p(x; \mu_{kj}, \Sigma_{kj}),\quad j=1,\ldots,J,
\end{equation}
 is considered as the emission distribution, then the following probabilities of the mixture components are computed in the E-step of the $(s+1)$th iteration of the EM algorithm
\begin{equation}\label{gamjk}
\gamma_{kj}^{(s+1)}(t) = \frac{\lambda_{kj}^{(s)}{\cal{N}}_p(x_t; \mu_{kj}^{(s)}, \Sigma_{kj}^{(s)})}{f_j(x_t) },
\end{equation}
where $\lambda_{kj}^{(s)}$, $\mu_{kj}^{(s)}$ and $\Sigma_{kj}^{(s)}$ are the $s$th updates of the emission parameters in the M-step of the $s$th iteration of the EM algorithm.

\subsection*{The M-step of the EM algorithm}
In the M-step of the EM algorithm, the initial probabilities are updated as follows
$$\pi_j^{(s+1)} = P(S_0 = j|X_{0:{\tau-1}} = x_{0:{\tau-1}};\theta^{(s)}) = L_j(0),
$$
For a semi-Markovian state $i$, the transition probabilities are updated as follows
 \begin{eqnarray}\label{18eq}
 p_{ij}^{(s+1)}&=& \frac{\sum_{t=0}^{\tau-2}G_j(t+1)p_{ij}^{(s)}F_i(t)}{\sum_{t=0}^{\tau-2}L1_i(t)}
 \end{eqnarray}
and for a Markovian state $i$
\begin{eqnarray}\label{15_eq}
\tilde{p}_{ij}^{(s+1)}&=& \frac{\sum_{t=0}^{\tau-2}\tilde{G}_j(t+1)\tilde{p}_{ij}^{(s)}\tilde{F}_i(t)}{\sum_{t=0}^{\tau-2}L_i(t)}
\end{eqnarray}
If we consider the mixture of multivariate normals, with the probability density function \eqref{mmn}, as the emission distribution,
then its parameters are updated as follows
$$ \lambda_{kj}^{(s+1)} = \frac{\sum_{t=0}^{\tau-1}\gamma_{kj}^{(s)}(t) L_j^{(s)}(t)}{\sum_{m=1}^{K_j}\sum_{t=0}^{\tau-1}\gamma_{mj}^{(s)}(t) L_j^{(s)}(t)} = \frac{\sum_{t=0}^{\tau-1}\gamma_{kj}^{(s)}(t) L_j^{(s)}(t)}{\sum_{t=0}^{\tau-1}L_j^{(s)}(t)},
$$$$\mu_{kj}^{(s+1)} = \frac{\sum_{t=0}^{\tau-1}\gamma_{kj}^{(s)}(t) L_j^{(s)}(t)x_t}{\sum_{t=0}^{\tau-1}\gamma_{kj}^{(s)}(t) L_j^{(s)}(t)},
$$$$
\Sigma_{kj}^{(s+1)} = \frac{\sum_{t=0}^{\tau-1}\gamma_{kj}^{(s)}(t) L_j^{(s)}(t)(x_t-\mu_{kj}^{(s+1)})(x_t-\mu_{kj}^{(s+1)})^T}{\sum_{t=0}^{\tau-1}\gamma_{kj}^{(s)}(t) L_j^{(s)}(t)}
$$
Also, the parameters of the sojourn time distribution are updated by maximization of the following quasi-log-likelihood function
$$\tilde{Q}_d(\{d_j(u)\}|\theta^{(s)}) = \sum_{u=1}^{M_j}\tilde{\eta}_{j,u}^{(s)}\log d_j(u),
$$
where
$$\tilde{\eta}_{j,u}=\sum_{t=0}^{\tau-2} G_j(t+1,u)\sum_{i\neq j} p_{ij} F_i(t) + A(u) d_j(u)\pi_j \prod_{v=1}^u \frac{f_j(x_{u_{\tau} - v)}}{N_{u_{\tau}-v}},$$
$$A(u) = \left\{\begin{array}{lr}\frac{L_{1j}(u-1)}{F_j(u-1)},& u\leq \tau-1\\ 1,& u>\tau-1,\end{array}\right.$$
and $u_{\tau} = \min(u,\tau)$.

\subsection*{Viterbi algorithm and smoothing  for the HHSMM model}
The Viterbi algorithm \citep{v67} is an algorithm to obtain the most likely state sequence, given the observations and the estimated parameters of the model.

For a semi-Markovian state $j$, and for $t=0,\ldots,\tau-2$, the probability of the most probable state sequence is obtained by the Viterbi recursion as follows
\begin{eqnarray}\label{24eq}
\alpha_j(t) &=& \max_{s_0,\ldots,s_{\tau-1}}P(S_{t+1}\neq j,S_t=j,S_{0:{t-1}}=s_{0:{t-1}},X_{0:t}=x_{0:t})\nonumber\\
&=& f_j(x_t)\max\Bigg[\max_{1\leq u\leq t}\Bigg[\Bigg\{\prod_{\nu=1}^{u-1}f_j(x_{t-\nu})\Bigg\}d_j(u)\max_{i\neq j}\{p_{ij}\alpha_i(t-u)\}\Bigg],\nonumber\\
&&\quad\times \Bigg\{\prod_{\nu=1}^{t}f_j(x_{t-\nu})\Bigg\}d_j(t+1)\pi_j\Bigg],
\end{eqnarray}
and
\begin{eqnarray}\label{25eq}
\alpha_j(\tau-1) &=& \max_{s_0,\ldots,s_{\tau-2}} P(S_{\tau-1} =j,S_{0:{\tau-2}} =s_{0:{\tau-2}},X_{0:{\tau-1}} =x_{0:{\tau-1}})\nonumber\\
&=&f_j(x_{\tau-1})\max\Bigg[\max_{1\leq u\leq \tau-1}\Bigg[\Bigg\{\prod_{\nu=1}^{u-1}f_j(x_{\tau-1-\nu})\Bigg\}D_j(u)\max_{i\neq j}\{p_{ij}\alpha_i(\tau-1-u)\}\Bigg],\nonumber\\
&&\quad\times \Bigg\{\prod_{\nu}^{\tau-1}f_j(x_{\tau-1-\nu})\Bigg\}D_j(\tau)\pi_j\Bigg],
\end{eqnarray}
For a Markovian state $j$, the Viterbi recursion is initialized by
$$\tilde{\alpha}_j(0) = P(S_0 = j,X_0 = x_0) = f_j(x_0)\pi_j$$
and for $t = 1,\ldots.\tau-1$,
\begin{eqnarray}\label{18_eq}
\tilde{\alpha}_j(t) &=& \max_{s_0,\ldots,s_{t-1}}P(S_t = j,S_0^{t-1} = s_0^{t-1},X_{0:t} = x_{0:t})\\
&=& f_j(x_t)\max_i\{\tilde{p}_{ij}\tilde{\alpha}_i(t-1)\}.
\end{eqnarray}
After obtaining the probability of the most probable state sequence, the current most likely state is obtained as $\hat{s}_t^* =\arg\max_{1\leq j\leq J} \alpha_j(t)$.

Another approach for obtaining the state sequence is the smoothing method, which uses the backward probabilities $L_j(t)$ instead of $\alpha_j(t)$.

\end{document}